\documentclass[fleqn,usenatbib]{mnras}

\usepackage[T1]{fontenc}

\DeclareRobustCommand{\VAN}[3]{#2}
\let\VANthebibliography\thebibliography
\def\thebibliography{\DeclareRobustCommand{\VAN}[3]{##3}\VANthebibliography}

\hypersetup{linkcolor=blue,citecolor=blue,filecolor=cyan,urlcolor=magenta}
\usepackage{graphicx}	
\usepackage{amsmath}	
\usepackage{amssymb}	
\usepackage{newtxtext,newtxmath}
\usepackage{color}
\usepackage{footmisc}
\usepackage{threeparttable}
\usepackage{hyperref}

\def\d{\mathrm{d}}
\def\mr{\mathrm}
\def\mc{\mathcal}
\def\nobs{n_{*,\rm obs}}
\def\fbeff{f_{\rm b,eff}}
\def\volrate{\mr{Gpc^{-3}\,yr^{-1}}}

\title[M81 FRB]{Implications of a rapidly varying FRB in a globular cluster of M81}

\author[Lu, Beniamini, \& Kumar]{
Wenbin Lu$^{1,2}$\thanks{wenbinlu@astro.princeton.edu}, Paz Beniamini$^{2,3}$\thanks{pazb@caltech.edu}, Pawan Kumar$^4$\thanks{pk@astro.as.utexas.edu}\\
  $^1$Department of Astrophysical Sciences, Princeton University, Princeton, NJ 08544, USA\\
  $^2$Walter Burke Institute for Theoretical Physics, Mail Code 350-17, Caltech, Pasadena, CA 91125, USA\\
  $^3$Astrophysics Research Center of the Open University (ARCO), The Open University of Israel, P.O Box 808, Ra’anana 43537, Israel\\
  $^4$Department of Astronomy, University of Texas at Austin, Austin, TX 78712, USA
}

\begin{document}
\label{firstpage}
\pagerange{\pageref{firstpage}--\pageref{lastpage}}
\maketitle

\begin{abstract}
A repeating source of fast radio bursts (FRBs) is recently discovered from a globular cluster of M81. Association to a globular cluster (or other old stellar systems) suggests that strongly magnetized neutron stars, which are the most likely objects responsible for FRBs, are born not only when young massive stars undergo core-collapse, but also by mergers of old white dwarfs. We find that the fractional contribution to the total FRB rate by old stellar populations is at least a few percent, and the precise fraction can be constrained by FRB searches in the directions of nearby galaxies, both star-forming and elliptical ones. Using very general arguments, we show that the activity time of the M81-FRB source is between 10$^4$ and 10$^6$ years, and more likely of order 10$^5$ years. The energetics of radio outbursts put a lower limit on the magnetic field strength of 10$^{13}\,$G, and the spin period $\gtrsim 0.2\,$s, thereby ruling out the source being a milli-second pulsar. The upper limit on the persistent X-ray luminosity (provided by \textit{Chandra}), together with the high FRB luminosity and frequent repetitions, severely constrains (or rules out) the possibility that the M81-FRB is a scaled-up version of giant pulses from Galactic pulsars. Finally, the 50 ns variability time of the FRB lightcurve suggests that the emission is produced in a compact region inside the neutron star magnetosphere, as it cannot be accounted for when the emission is at distances $\gtrsim 10^{10}\rm\, cm$.
\end{abstract}




\vspace{0.7cm}

\section{Introduction}

Fast radio bursts (FRBs) are bright, short-duration radio pulses \citep{Lorimer+07, Thornton+13} whose origins are so far unclear. Most FRBs discovered to date are generally believed to come from cosmological distances as suggested by the dispersion measure-redshift (DM-z) relation \citep{macquart20_DM_z_relation}. The recent discovery of FRB200428 from an otherwise unremarkable Galactic soft gamma repeater \citep[SGR,][]{bochenek20_FRB200428, chime20_FRB200428, Lin+20, Ridania2020, Li2020, Mereghetti+20, Tavani2020} provided a number of important clues: \citep[see][]{LKZ2020, Margalit+20} (1) at least some FRBs are produced by highly magnetized neutron stars --- magnetars \citep{katz82_magnetar, thompson95_magnetars}, (2) these magnetars undergo sudden dissipation of magnetic energy most of which is converted into X-ray emission while a small fraction is channelled into coherent radio emission, (3) the entire FRB population roughly follows a power-law rate density function that extends from an isotropic-equivalent specific energy of $\sim 10^{33}\rm \,erg\,Hz^{-1}$ (the brightest cosmological ones) down to $\sim 10^{26}\rm\, erg\,Hz^{-1}$ (that of the Galactic FRB), and (4) some sources are much more active at repeatedly generating FRBs than others. 

The spatial distribution of SGRs (and anomalous X-ray pulsars or AXPs) in the Milky Way being within a small scale-height $\lesssim30\rm\, pc$ from the Galactic mid-plane and some of them being associated with young supernova remnants indicate that most of them are recently born from the core-collapse of massive stars \citep{kaspi17_magnetars}. Many other FRBs with precise localizations are found to be associated with late-type star-forming galaxies \citep{Chatterjee+17, Tendulkar+17, marcote20_R3_localization, ravi19_DSA_first_localization, Prochaska+19, Bannister+19, heintz20_host_population, law20_FRB_realfast} \citep[see][for another recently discovered star-forming host galaxy]{ravi21_FRB201124_host_gal, fong21_FRB201124_host}. These host localizations have been used to argue against the hypothesis that the FRB rate strictly follows the distribution of stellar mass \citep{li20_FRB_host_comparison, heintz20_host_population, bochenek21_FRB_host_comparison}. However, formation models that follow the star-formation rate \citep{james21_FRB_population_evolution} or a combination of star-formation rate and stellar mass \citep[e.g., compact object mergers,][]{margalit19_merger_formation, wang20_bns_mergers} are not ruled out.

More recently, the discovery of FRB20200120E localized to a globular cluster of the very nearby M81 galaxy \citep[hereafter M81-FRB,][]{CHIMEM81, KirstenM81} raises a number of further questions: What is the nature of the source? What is its formation pathway and how it may differ from other FRB sources? What fraction of the cosmological FRB rate is contributed by old vs. young stellar populations? This paper aims to answer these questions by considering the M81-FRB in the broader FRB source population. Furthermore, high time resolution observations of the brightest bursts from the M81-FRB source showed that the light curves had rapid variability on timescales of $50\rm\, ns$ \citep[or less, as limited by the signal-to-noise ratio in each time bin,][]{nimmo21_short_var_timescale, MajidM81}, and we aim to provide constraints on the emission mechanism of the coherent radio waves.

This paper is organized as follows. In \S \ref{sec:n_star}, we estimate the number density of the M81-FRB-like sources and other subclasses of FRBs and then discuss the implications on their formation pathways. Then, in \S \ref{sec:luminosity_density}, we calculate the time-averaged luminosity density --- the product of time-averaged luminosity and number density, contributed by each subclass of FRB sources and then discuss the contributions to the total FRB rate by young vs. old stellar populations.  In \S \ref{sec:M81-FRB-energetics}, we combine all the arguments to constrain the nature of the M81-FRB source --- it is consistent with a slowly spinning NS with strong B-fields ($B\gtrsim 10^{13}\rm\, G$) formed by a double white dwarf merger. In \S \ref{sec:variability-time}, we show that the rapid variability in the light curves of the M81-FRB cannot be generated by propagation effects far away from the emitting plasma and is hence intrinsic to the emission process. Some selection effects that favor the identification of a globular cluster-hosted source over others in the galactic field of the M81 galaxy are briefly discussed in \S \ref{sec:selection_effects}. We summarize our main results in \S \ref{sec:summary}.  

Throughout this paper, we make the assumption that FRBs are generated by neutron stars (NSs, but not necessarily like the Galactic SGRs), as supported by observations of the Galactic FRB200428 \citep{bochenek20_FRB200428} as well as many theoretical arguments \citep{LuKumar2018}. The readers are referred to \citet{waxman17_accretion_powered, katz17_accretion_powered, sridhar21_accretion_powered_FRB} who suggested that FRBs may be powered by accretion (and NSs are not necessary). We use the convenient subscript notation of $X_n \equiv X/10^n$ in the CGS units.

\section{Source number density of FRB subclasses}\label{sec:n_star}
The initial discovery of the first repeating FRB source \citep[FRB 20121102A, hereafter R1,][]{spitler16_FRB121102, scholz16_FRB121102} sparked discussion whether all other sources (mostly found by the Parkes telescope at that time) repeat as frequently as R1 does. It was clear that the number density of R1-like sources must be less than a few $\times10^3\rm\, Gpc^{-3}$ \citep{lu16_universalEDF} to avoid over-predicting the all-sky FRB rate as inferred from the detections by Parkes telescope \citep{keane15_ParkesFRBrate, rane16_ParkesFRBrate,champion16_ParkesFRBrate}. This means that R1-like sources are very rare in the Universe, with a birth rate less than $10^{-3}$ of the core-collapse supernova (ccSN) rate, provided that the lifetime for the repeating activity is longer than 10 years \citep{lu16_universalEDF}.

Subsequent follow-up observations of many apparently non-repeating FRBs show that, in fact, most of them repeat much less frequently (above a given energy/luminosity threshold) than R1 and hence their source number density could be higher \citep[e.g.,][]{james20_frb_repeating_rate}. This conclusion became more robust thanks to the CHIME/FRB survey \citep{chime21_catalog1} which unveiled that only a small fraction of FRB sources show repeated bursts above the telescope's detection threshold --- the repeater fraction (as well as the DM distribution of repeaters) strongly depends on the distribution of repeating rates among the source population\footnote{For instance, the first CHIME catalog \citep{chime21_catalog1} includes 30 single sources with DM excess of less than $150\rm\, pc\, cm^{-3}$ beyond the Galactic interstellar medium (ISM) contribution. If one conservatively removes the DM contribution of $\sim$$50\rm\, pc\,cm^{-3}$ from the circum-galactic medium of the Milky Way as well as the ISM and halo of the host galaxy \citep[e.g.,][]{shull18_CGM_IGM_DM, prochaska19_haloDM, keating20_haloDM}, these sources are expected to be within redshift of 0.1 according to the Macquart relation \citep{macquart20_DM_z_relation}, so the CHIME survey is sufficiently sensitive to rule out a repeating rate comparable to or higher than that of R1 for most of these 30 sources.} \citep{lu20_CHIME_FRB_population}. Another independent piece of information is that FRB20180916B \citep[hereafter R3,][]{chime19_FRB180916, marcote20_R3_localization}, which has an average repeating rate of a few per day above specific energy of $10^{29}\rm\, erg\,Hz^{-1}$ (slightly lower than R1's repeating rate), is the only source with such a high repeating rate within its distance of 150 Mpc in the northern half of the sky. We conclude that the highly active R1/R3-like objects are only a minority of FRB sources in terms of number density, although they could still contribute a large fraction of the total volumetric rate due to their high repeating rates. It is so far unclear whether the cosmological FRB rate is dominated by the most actively repeating sources or those that rarely repeat.

\begin{figure*}
\centering
\includegraphics[width = 0.7\textwidth]{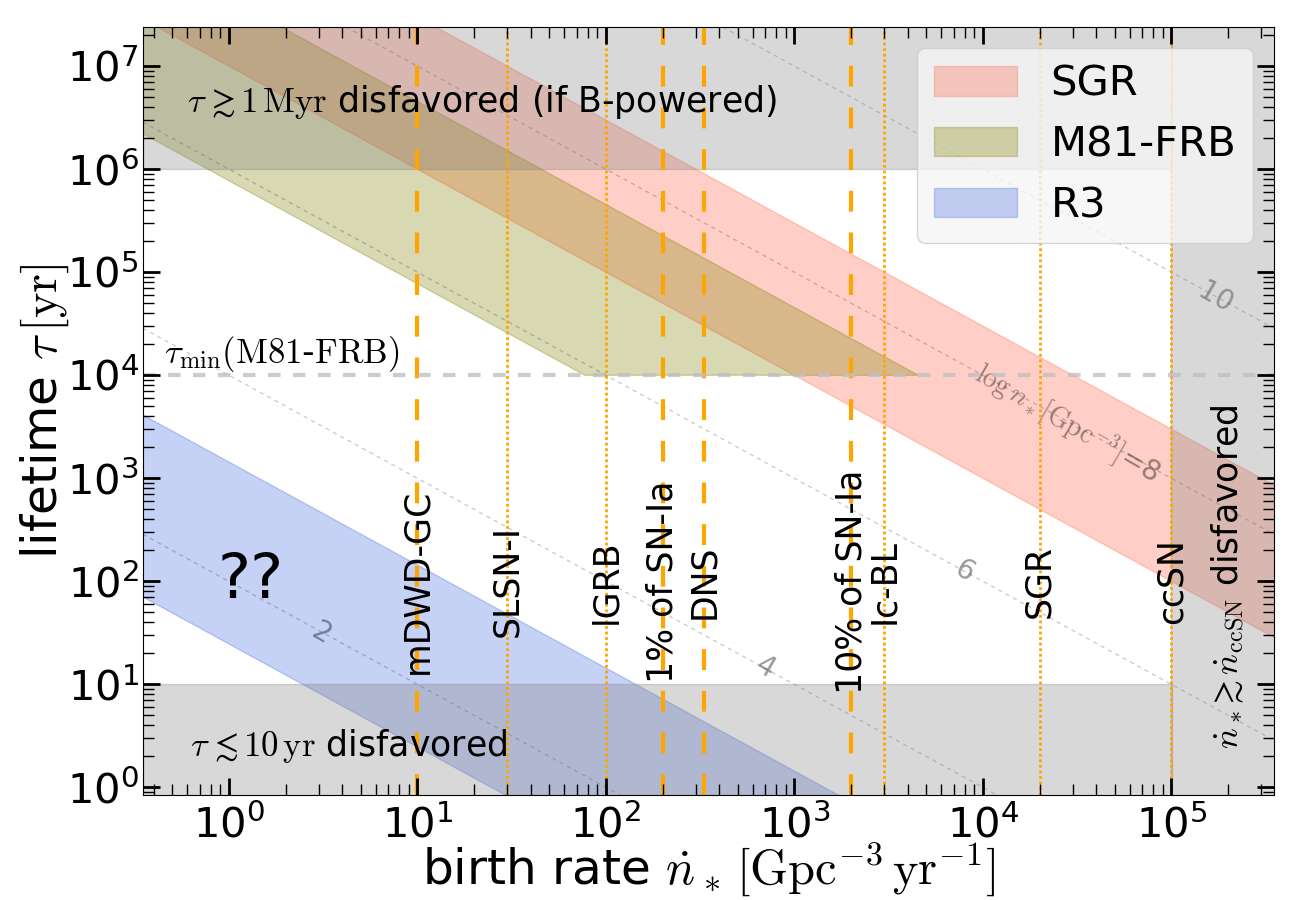}
\caption{Constraints on the lifetime $\tau$ and birth rate $\dot{n}_*$ of each of the three subclasses of FRB sources we have considered: SGR (red), M81-FRB (green), and R3 (blue). Each colored band represents the allowed (90\% CL) source number density of each subclass. We adopt a beaming correction of $f_{\rm b,eff}=0.3$ for the M81-FRB and R3 subclasses of sources. The estimated birth rates of various potential compact-object formation channels in the local Universe are marked by orange vertical lines: massive (super-Chandrasekhar) double white dwarf mergers in core-collapsed globular clusters \citep[mDWD-GC, $\sim 10\rm\, Gpc^{-3}\, yr^{-1}$,][]{kremer21_massive_WD_mergers}, type-I superluminous supernovae \citep[SLSN-I, $\sim 30\rm\, Gpc^{-3}\, yr^{-1}$,][]{quimby13_SLSN-I_rate, prajs17_SLSN-I_rate, frohmaier21_sn_rates}, long gamma-ray bursts \citep[lGRB, $\sim 100\rm\, Gpc^{-3}\, yr^{-1}$,][]{wanderman10_lGRB_rate, sun15_grb_rates}, double neutron star mergers \citep[DNS, $\sim 300\rm\, Gpc^{-3}\, yr^{-1}$,][]{LVC21_GW_population_rates}, 1\% of the type-Ia supernova rate \citep[1\% of SN-Ia, $\sim 200\rm\, Gpc^{-3}\, yr^{-1}$,][]{frohmaier19_typeIa_rate}, broad-lined type Ic supernovae \citep[Ic-BL, $\sim 3\times10^3\rm\, Gpc^{-3}\, yr^{-1}$,][]{shivvers17_Ic-BL_rate}, birth rate of soft gamma repeaters \citep[SGR, $\sim2\times10^{4}\rm\, Gpc^{-3}\, yr^{-1}$,][]{beniamini19_magnetar_birth_rate}, core-collapse supernovae \citep[ccSN, $\sim10^5\rm\, Gpc^{-3}\,yr^{-1}$,][]{li11_supernova_rates}. We note that some of these rates are only rough estimates that are accurate to within an order of magnitude. The dotted (or dashed) lines are for the channels that (do not) require young stellar population. The minimum lifetime of the M81-FRB source is constrained to be $\tau_{\rm min}(\mbox{M81-FRB}) \sim 10^4\rm\, yr$ (which is also the lifetime of typical Galactic magnetars) and is marked by a silver horizontal dashed line. The grey-dotted lines are the contours for the logarithmic number density, $\mr{log}\,n_*[\rm Gpc^{-3}]$. The region with $\tau\lesssim 10\rm\, yr$ is disfavored for many observed repeaters which show non-decaying activity for many years \citep[and the expanding supernova remnant may produce a time-dependent DM that is inconsistent with observations,][]{PG2018}. Very long source lifetime of $\tau\gtrsim 1\rm\, Myr$ is disfavored if FRB emission is powered by free magnetic energy stored in the NS interior.  The region with $\dot{n}_*\gtrsim 10^5\rm\, Gpc^{-3}\, yr^{-1}$ is also disfavored because the birth rate of neutron stars and black holes cannot greatly exceed the rate of ccSNe. The parameter space with $\dot{n}_*\lesssim 10\rm\, Gpc^{-3}\,yr^{-1}$ is so far largely unexplored by electromagnetic transient surveys. }
\label{fig:source-density}
\end{figure*}

In this section, we provide robust constraints on the number density of each of the subclasses of FRBs and discuss the implications. 

The volumetric number density $\nobs$ of a given class of objects can be statistically constrained by the distance $D_1$ of its closest member to the observer, or equivalently the enclosed volume $V_1$ within $D_1$. We note that the number density $\nobs$ is defined based on the \textit{observable} sources, whereas an additional beaming factor is needed to account for the ones whose FRBs are permanently beamed away from us so as to recover the true number density
\begin{equation}
    n_* = \nobs/\fbeff,
\end{equation}
and throughout this work, we adopt a conservative fiducial beaming factor\footnote{Due to the spin of the source object, the effective beaming fraction $f_{\rm b,eff}$, defined as the total solid angle covered by all bursts from this source divided by $4\pi$, is likely not much smaller than 0.1 (unless the spin axis is closely aligned with the FRB beams). The true source number density is likely a factor of $\sim10$ larger than that of the observable sources. A very conservative lower limit on $f_{\rm b,eff}$ can be obtained by the fact that the two sub-pulses from FRB200428 were separated by a duration of $0.01P$ ($P$ being the spin period), so this constrains $f_{\rm b,eff}\gtrsim 0.01$. Empirically, the fact that we have detected FRB200428 from one of the $\sim$30 magnetars in our Galaxy suggests $f_{\rm b,eff}\gtrsim 1/30$ at least for SGR-like sources.} of $f_{\rm b,eff}=0.3$, but our results can be easily scaled to other choices of beaming factor once more information is available.

For a given number density $\nobs$, the probability density function of the enclosed volume $V_1$ of the closest member is
\begin{equation}
    \left.{\d P\over \d V_1}\right|_{\nobs} = f(V_1, \nobs) = \mr{e}^{-V_1 \nobs} \nobs,
\end{equation}
where the first factor $\mr{e}^{-V_1 \nobs}$ is the Poisson probability of having zero source within a volume of $V_1$ and the second factor $\nobs\d V_1$ is the probability of having one source in the volume of $\d V_1$.  Taking a non-informative prior distribution of $\d P_0/\d \nobs = \nobs^{-1}$ \citep[i.e., Jeffery's prior,][]{damien2013bayesian}, we obtain the posterior distribution of the source number density $\nobs$ according to the Bayes Theorem
\begin{equation}\label{eq:number_density_posterior}
    {\d P \over \d \nobs} = V_1f(V_1, \nobs) \nobs^{-1} = V_1 \mr{e}^{-V_1 \nobs}.
\end{equation}
This can be converted into a cumulative distribution $P(<\nobs) = 1 - \mr{exp}(-V_1 \nobs)$, which means that the 90\% or 68\% confidence interval (CL) is $\nobs\in(0.051, 3.0)V_1^{-1}$ or $(0.17, 1.8)V_1^{-1}$, respectively. We note that, although the median value is $\approx 0.7V_1^{-1}$, there is non-negligible probability that $\nobs\ll V_1^{-1}$ (whereas the probability for $\nobs\gg V_1^{-1}$ is exponentially suppressed).


Let us now consider two different classes of sources that are M81-FRB-like (which are unique for their globular cluster environment) and R3-like (unique for their prolific repeating activities). The first class has its closest member at $\tilde{D}_1=3.6\rm\,Mpc$ \citep{KirstenM81} and $\tilde{V}_1\simeq 2\pi \tilde D_1^3/3 = 98\mr{\,Mpc^{3}}$ (note that CHIME covers roughly the northern half of the sky). However, since the number density of galaxies in the very local Universe ($\lesssim 5\rm\, Mpc$) is enhanced compared to the average, it is important to correct for this bias. Based on the stellar mass of M81 \citep[about $7\times10^{10}M_\odot$,][]{deblok08_M81_stellar_mass} and the cosmic stellar mass density of $6\times10^{8}M_\odot\rm\,Mpc^{-3}$ \citep{madau14_cosmic_SFR}, we infer the stellar-mass weighted average volume for M81-FRB-like sources to be $V_1(\mbox{M81-FRB})\simeq 220\mr{\, Mpc}^3$, so the number density is in the range $2.3\times10^5 < \nobs(\mbox{M81-FRB}) < 1.4\times10^7\rm\, Gpc^{-3}$ at 90\% confidence level (CL). The second class has its closest member at $D_1=149\rm\, Mpc$ \citep{marcote20_R3_localization} and $V_1\simeq 2\pi D_1^3/3 \simeq 6.9\times10^6\rm\, Mpc^{3}$, so the number density is in the range $7.4<\nobs(\mbox{R3})<432\rm\, Gpc^{-3}$ (90\% CL). 
We note that the uncertainties in $\nobs$ are dominated by Poisson fluctuation, so taking a more precisely determined exposure-weighted sky area coverage for the CHIME survey would only lead to a small change in the constraints we have obtained.


SGRs belong to another class of FRB sources, and a prototype is the Galactic magnetar SGR 1935+2154 which generated FRB200428 as observed by STARE2 and CHIME \citep{bochenek20_FRB200428, chime20_FRB200428}. Since the cosmic volume shared by the Milky Way is\footnote{This is obtained by taking either the star formation rate(SFR)-weighted or the stellar mass-weighted average volume. Using the Galactic SFR and stellar mass given by \citet{licquia15_Galactic_SFR_stellar_mass} and the cosmic SFR density and stellar mass density in the local Universe given by \citet{madau14_cosmic_SFR}, we find that the results from these two estimates are similar.
} $\simeq 100\rm\, Mpc^{3}$, the number density of \textit{FRB-producing} SGR sources is in the range $n_*\in (10^7, 3\times10^8)\rm\, Gpc^{-3}$, where the upper limit is obtained by assuming all $\sim30$ Galactic SGRs produce FRBs at an equal rate.

Let us denote the lifetime of each subclass of FRB sources as $\tau$, during which the average activity level stays roughly constant and after which they become much less active. Under the steady-state assumption, we can infer the birth rate to be
\begin{equation}
    \dot{n}_*= n_*/\tau.
\end{equation}
In Fig. \ref{fig:source-density}, we show the constraints on the lifetime and birth rate of each of the three subclasses of FRB sources, along with the estimated rates of many classes of transients. Based on empirical knowledge, we disfavor the regions with lifetime $\tau\lesssim 10\rm\, yr$ because: (1) some sources (e.g., R1 and many CHIME repeaters) have been known to repeat for more than a few years; and (2) the compact object was presumably born in an explosion which was accompanied by an ejecta of mass in the range $10^{-1}$--$100\rm\, M_\odot$ and the DM contribution by parts of the ejecta that are heated by the shocks (due to interactions with the pulsar wind nebula or the circum-stellar medium) or photo-ionization may produce a time-dependent DM with noticeable evolution in the first decade \citep[this is inconsistent with weakly- or non-evolving DM in many sources,][]{yang17_DM_evolution, PG2018, margalit19_merger_formation}. We also disfavor the regions with source birth rate greater than $10^{5}\rm\, Gpc^{-3}\, yr^{-1}$, which is the ccSN rate.


In the following, we discuss the potential formation pathways of each class of sources.


The lifetime and birth rate of M81-FRB-like sources are only loosely constrained due to their degeneracy. A model-independent way of constraining the source lifetime is to assume that all globular-cluster NSs are capable of generating FRBs in a fraction of the cluster's age. The total number of NSs in all Galactic globular clusters is estimated to be of the order $10^5$ \citep{pfalh02_GC_NSs, ivanova08_GC_NSs, kremer20_CMC_catalog}, and this gives a cosmic number density of $\sim10^{12}\rm\, Gpc^{-3}$. If only a fraction $\tau/(10\rm\, Gyr)$ of these NSs are generating FRBs at a given moment, the number density of currently active FRB sources is $\sim 10^6(\tau/10^4\mr{\,yr}) \rm\, Gpc^{-3}$. This, combined with the number density constraints $n_*(\mbox{M81-FRB}) = f_{\rm b,eff}^{-1}\nobs \in (7.7\times10^5, 4.5\times10^7)(f_{\rm b,eff}/0.3)^{-1}\rm\, Gpc^{-3}$ (90\% CL), requires the lifetime of the M81-FRB to be $\tau \gtrsim 10^4(f_{\rm b,eff}/0.3)^{-1}\rm\, yr$.


It has been proposed that mergers of white dwarfs (WDs) with total mass exceeding the Chandrasekhar mass may give rise to low-mass NSs \citep{shen12_WD_merger_remnants, schwab16_massive_WD_mergers, schwab21_WD_merger_remnant}, although the final fate of the remnant (either a massive WD or low-mass NS) strongly depends on the highly uncertain mass loss rate during the 10 kyr red giant evolutionary phase after the merger. Rapidly rotating, magnetized NSs may also be generated from the accretion-induced collapse (AIC) channel where an O-Ne-Mg WD accretes gas from a companion star and grows past the Chandrasekhar limit \citep[e.g.,][]{tauris13_AIC_binary_accretion}, although the outcome (either AIC or SN-Ia explosion) sensitively depends on the uncertain competition between oxygen deflagration and electron captures onto Ne and Mg \citep{nomoto91_AIC_conditions}. The total mass of globular clusters in the Milky Way is $3\times 10^{7} M_\odot$ \citep{baumgardt18_MW_GC_catalog} and roughly $\sim 10\%$ of the mass is in massive ($\sim1M_\odot$) WDs \citep{kremer20_CMC_catalog, rui21_CMC_matching_observations}. This yields total mass density of $3\times 10^{13}\rm\, M_\odot\, Gpc^{-3}$. In the very extreme case where half of these massive WDs turn into young magnetized NSs within the globular cluster lifetime of 10 Gyr, this gives an upper limit for the NS birth rate $\dot{n}_*<1.5\times 10^{3}\rm\, Gpc^{-3}\, yr^{-1}$ (i.e., less than 10\% of the SN-Ia rate) from this channel. This birth rate upper limit can be combined with the number density constraint to yield a lower limit for the lifetime for the M81-FRB source $\tau>10^3(f_{\rm b,eff}/0.3)^{-1}\rm\, yr$ (which is extremely conservative).


A recent study by \citet{kremer21_massive_WD_mergers} showed that dynamical interactions in the innermost regions of core-collapsed globular clusters, where the mass density is dominated by massive ($\sim1M_\odot$) WDs, lead to super-Chandrasekhar double WD mergers at a rate $\sim10\rm\, Gpc^{-3}\,yr^{-1}$. This means that only a very small fraction of the globular-cluster WDs may turn into NSs. If these mergers indeed produce young magnetized NSs as M81-FRB-like sources, then a birth rate of $10\rm\, Gpc^{-3}\,yr^{-1}$ implies that the source lifetime is $\tau  \in (8\times10^4, 5\times 10^6)\times (f_{\rm b,eff}/0.3)^{-1}\rm\, yr$.

In Appendix \ref{sec:Bfield-decay-time}, we show that the upper limit on the NS magnetic activity lifetime is $\tau\lesssim 1\rm\, Myr$. This sets a lower limit for the birth rate to be $\dot{n}_*(\mbox{M81-FRB})\gtrsim 1 (f_{\rm b,eff}/0.3)^{-1}\rm\, Gpc^{-3}\, yr^{-1}$, where the $f_{\rm b,eff}^{-1}$ factor accounts for the unseen sources whose FRBs are permanently beamed away from us. This does not rule out the hypothesis that the dynamically assembled super-Chandrasekhar WD mergers in globular clusters produce FRB-emitting NSs at a rate of the order $10\, \volrate$ \citep{kremer21_massive_WD_mergers}.

The current constraints also leave open the possibility that FRB-emitting NSs are produced by super-Chandrasekhar WD mergers \citep{shen12_WD_merger_remnants, schwab16_massive_WD_mergers} or AIC \citep{margalit19_merger_formation} at a comparable or higher rate from old stellar populations in the \textit{galactic field} than in globular clusters. This is because the stellar mass contained in the Galactic thick disk and halo is a few hundred times higher than the mass in globular clusters \citep{robin03_galactic_stellar_structures, deason19_MW_halo_mass}. From Fig. \ref{fig:source-density}, we see that the available constraints on the M81-FRB subclass indeed allow a source birth rate in the range $(1, 10^{3.5})\times (f_{\rm b,eff}/0.3)^{-1}\rm\, Gpc^{-3}\, yr^{-1}$, the upper limit being close to 10\% of the SN-Ia rate. Statistical studies of the Galactic population of binary WD systems find a super-Chandrasekhar merger rate of the order of 10\% of the SN-Ia rate \citep[][]{badenes12_WD_merger_rate}, and it is likely that a fraction of these mergers generate strongly magnetized NSs \citep[see][]{caiazzo2021_magnetized_massive_WD}. In the next section, we show that the fractional contribution to the total FRB rate by old stellar populations is at least a few percent, and hence we predict that future interferometric observations of cosmological FRBs will find a fraction of at least a few percent of them either in elliptical galaxies or at large spatial offsets from the centers of the host galaxies.

Finally, we turn our attention to R3-like sources. We see from Fig. \ref{fig:source-density} that R3-like sources must have an extremely low birth rate of $\dot{n}_*(\mr{R3}) \lesssim 100 (f_{\rm b,eff}/0.3)^{-1} \rm\, Gpc^{-3}\,yr^{-1}$. This is in agreement with the earlier results by \citet{lu16_universalEDF} that the number density of R1/R3-like sources must be small such that their birth rate is less than $10^{-3}$ of the ccSN rate\footnote{It has been speculated that some very rare magnetars (that are not probed by the Galactic magnetar population) with extremely strong B-fields $>10^{16}\rm\, G$ may be responsible for the R1/R3-like sources \citep{Metzger+17, margalit19_merger_formation}. It has also been proposed that the 16-day periodicity of R3 \citep{chime20_R3_periodicity} may be the mark of an ultra-long spin period, arising from such an extreme magnetar \citep{Beniamini+20}. } \citep[see also][who reached the same conclusion]{nicholl17_source_number_density}.
Current optical transient searches/classifications \citep{perley20_ZTF_transient_demographics} have not been able to constrain the demographics of rare explosions whose event rates are of the order $10\rm\, Gpc^{-3}\, yr^{-1}$, especially the ones much fainter than normal type II supernovae. The next generation of transient surveys, such as the Vera Rubin Observatory \citep[][]{Ivezic19_LSST}, will increase the current number of transients by an order of magnitude, and it may be feasible to associate R3-like sources to their birth events (if they are optically bright). We also suggest that the lifetime of R3-like sources may be less than 1 kyr, provided that the birth rate of their sources is higher than $1\rm\, Gpc^{-3}\, yr^{-1}$ (an extremely small birth rate may require fine-tuning in modeling the formation pathway).

\section{Time-averaged Luminosity density}\label{sec:luminosity_density}

In the previous section, we have discussed the number densities and birth rates of the three subclasses of FRB sources: SGR, M81-FRB, and R3. An interesting trend is that the sources with lower number densities have higher repeating rates. It is so far unclear which subclass dominates the cosmological FRB rate. This question will be addressed in this section.

The CHIME repeater catalog\footnote{\href{https://www.chime-frb.ca/repeaters}{https://www.chime-frb.ca/repeaters}} has recorded 7 bursts from the M81-FRB and 73 bursts from R3 at the time of writing, and these two sources have very similar declinations of 66$^\circ$ and 69$^{\circ}$ (meaning that the total exposure times are similar), so the time-averaged radio luminosity in FRB bursts from these two sources differ by a factor of $(73/7)\times (149\mr{Mpc}/3.6\mr{Mpc})^2\simeq 1.8\times10^4$. On the other hand, the enclosed volume of the closest members in these two subclasses of sources differ by a factor of $3\times10^4$. A surprising result is that the luminosity density, which is the product of \textit{observable} source number density (without beaming correction) and time-averaged isotropic equivalent luminosity, of these two distinct classes of sources are similar to each other. The most active R3-like (but rare) repeaters may contribute a similar luminosity density as the much less active M81-FRB-like (but abundant) ones.

The average fluence of bursts detected from M81-FRB and R3 is about $2\rm\, Jy\, ms$ with a small uncertainty of about a factor of two (which can in principle be eliminated once we know the telescope's response at the moment of each detection). The average isotropic equivalent energies ($\nu E_\nu$ for $\nu = 600\rm\, MHz$) of the detected bursts from M81-FRB and R3 are $E_{\rm frb, M81}\sim 2\times10^{34}\rm\, erg$ and $E_{\rm frb, R3} \sim 3\times10^{37}\rm\, erg$. Given that the on-source time for these two sources are both about $100 \rm\, hr$ \citep{chime21_catalog1}, we infer the time-averaged isotropic equivalent luminosity\footnote{Note that the \textit{observed} time-averaged luminosity (based on detected bursts) is only a lower limit for the total luminosity of the source, because fainter ones below the fluence threshold and brighter ones occurring at a very low rate are missed by the survey. Given the lack of statistical analysis of the repeating rate as a function of burst fluences for these two sources, we are unable to do an accurate incompleteness correction in this work.  However, studies of the entire FRB population \citep{LuPiro19_ASKAP_population, luo20_FRB_luminosity_function, lu20_CHIME_FRB_population, james21_FRB_population_evolution} showed that the volumetric rate density is consistent with each individual source repeat with an power-law energy distribution of $\mc{R}_{\rm rep}(>E_{\rm frb})\propto E_{\rm frb}^{-\alpha}$ with $\alpha\simeq 1$, implying that the incompleteness correction factor is likely of order unity. \label{fn:alpha}} of these two sources to be $\bar{L}_{\rm M81}\sim 3\times10^{29}\rm\, erg\, s^{-1}$ and $\bar{L}_{\rm R3} \sim 6\times10^{33}\rm\, erg\,s^{-1}$. We then combine the time-averaged luminosities with the number densities of these two classes of sources and obtain their contributions to the cosmological luminosity density
\begin{equation}\label{eq:mcL_M81FRB}
    \mc{L}(\mbox{M81-FRB})\equiv \nobs\bar{L}\in (7\times10^{34}, 4\times10^{36})\rm\, erg\, s^{-1}\, Gpc^{-3},
\end{equation}
and
\begin{equation}\label{eq:mcL_R3}
    \mc{L}(\mbox{R3})\equiv \nobs\bar{L}(\mbox{R3})\in (5\times10^{34}, 3\times10^{36})\rm\, erg\, s^{-1}\, Gpc^{-3}.
\end{equation}

The FRB luminosity density contributed by the SGR class of sources can be estimated as follows.

The STARE2 survey \citep{bochenek20_STARE2_instrument, bochenek20_FRB200428} has a field of view with full-width half-maximum of about 70$^{\rm o}$ and is sensitive to Galactic FRBs with fluences of the order of $\rm MJy\, ms$. In the time-averaged sense, it covers about 1/20 of the volume of the Milky Way's disk \citep[see][]{connor21_GReX}. Over an operational time of about 2 years, STARE2 detected one burst, FRB 200428, which had a fluence of $1.5\rm\, MJy\,ms$ and isotropic equivalent energy ($\nu E_\nu$ for $\nu=1.4\rm\, GHz$) of $E_{\rm frb, 200428}\sim 2.5\times 10^{35}\rm\, erg$ for a distance of 10 kpc \citep{bochenek20_FRB200428}. To account for for Poisson fluctuations, we infer the expectation number $\lambda$ from the single detection using the Bayes Theorem
\begin{equation}
    {\d P\over \d \lambda} = {2\over \sqrt{\pi}} \lambda \mr{e}^{-\lambda} \times \lambda^{-1/2},
\end{equation}
where the $\lambda^{-1/2}$ factor is the non-informative Jeffery's prior for Poisson distribution \citep{damien2013bayesian} and the normalization factor of $2/\sqrt{\pi}=[\Gamma(3/2)]^{-1}$ is given by the Gamma function $\Gamma(x)$. Thus, the expectation number of detections by STARE2 is in the range $\lambda\in(0.18, 3.9)$ at 90\% CL with a median of $\approx 1.2$, and hence the event rate above the fluence of FRB 200428 in the entire Milky Way is given by $\mc{R}_{\rm MW}\in (1.8, 39)\rm\, yr^{-1}$ at 90\% CL with a median of $12\rm\, yr^{-1}$.

Then, we use the cosmic volume shared by the Milky Way, which is $\simeq 100\rm\, Mpc^{3}$, to obtain the FRB luminosity density contributed by the SGR class of objects
\begin{equation}
    \mc{L}(\mbox{SGR})\equiv \nobs\bar{L}(\mbox{SGR})\in (1\times10^{35}, 3\times10^{36})\rm\, erg\,s^{-1}\,Gpc^{-3}.
\end{equation}
Note that this is the measurement of the product of the observable number density $\nobs$ and mean luminosity $\bar{L}(\mr{SGR})$ by radio observations only, whereas the individual quantities are not well measured. Here, we argue that $\bar{L}(\mr{SGR})$ is in the range $(5\times10^{26}, 10^{28})\rm\, erg\,s^{-1}$, where the lower limit is obtained by taking the extreme values of $\mc{L}(\mbox{SGR}) = 1.4\times10^{35}\rm\, erg\, s^{-1}\, Gpc^{-3}$ and $\nobs = 3\times10^8\rm\, Gpc^{-3}$ (assuming all $\sim30$ Galactic magnetars produce observable FRBs at an equal rate), and the upper limit comes from the non-detection of repetition from SGR 1935+2154 above the STARE2 threshold in about 0.4 yr of on-source time. Thus, we take $\bar{L}(\mr{SGR})\sim 2\times10^{27}\rm\, erg\,s^{-1}$ as a rough estimate of the median value (this is not crucial to our conclusions).

\begin{figure}
\centering
\includegraphics[width = 0.48\textwidth]{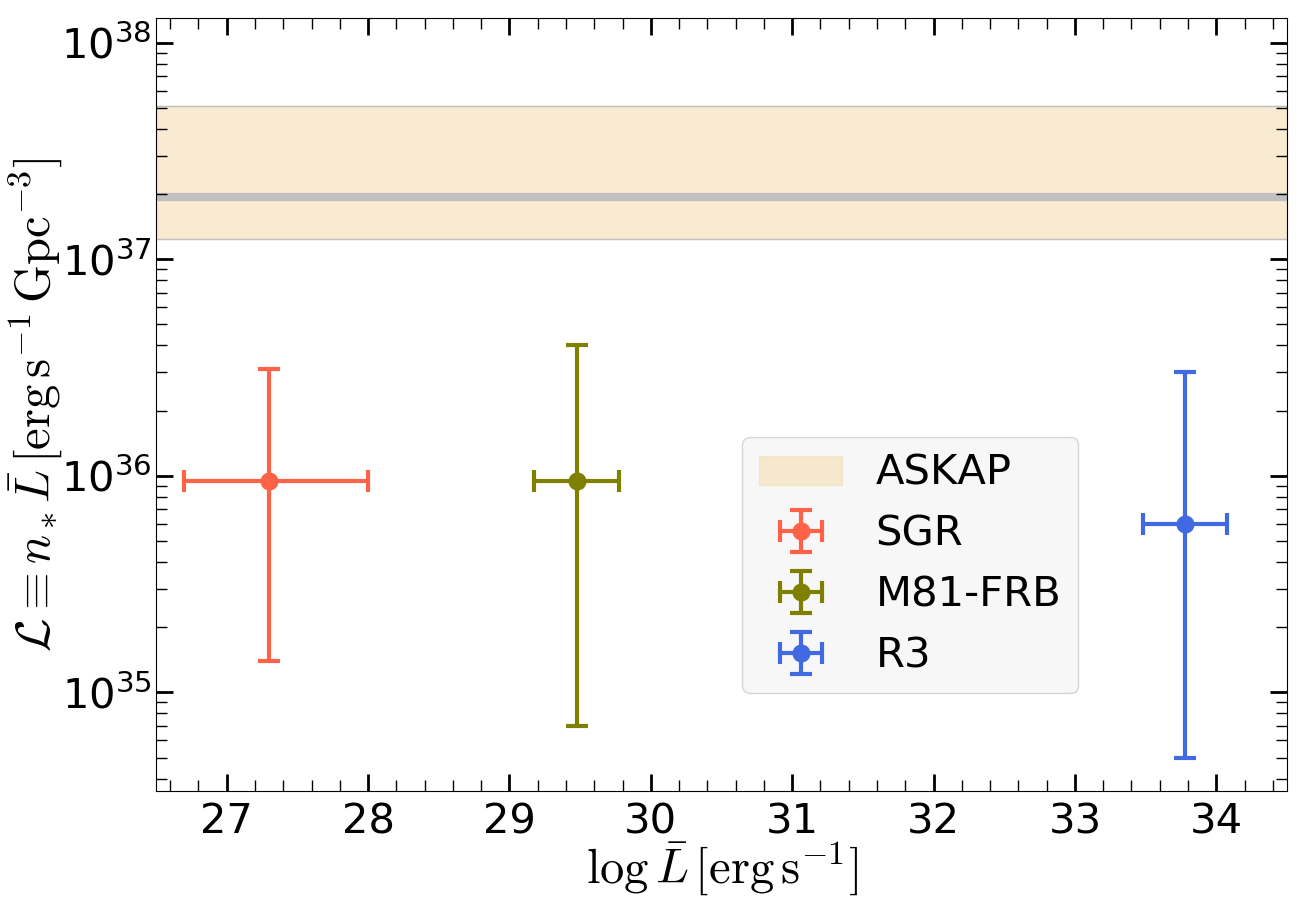}
\caption{The vertical axis shows the luminosity density contributed by each of the three classes of FRB sources SGR (red), M81-FRB (green), and R3 (blue), and the errors reflect the Poisson uncertainties ($90\%$ CL) of the source number densities. The horizontal axis shows the time-averaged isotropic equivalent luminosity of each of the sources, and the errors for $\bar{L}\mbox{(M81-FRB)}$ and $\bar{L}(\mbox{R3})$ are taken to be $0.3$ in log space. The shaded region shows the luminosity density (90\% CL) contributed by all bursts above a minimum specific energy of $10^{31}\rm\, erg\,Hz^{-1}$ as constrained by the ASKAP-detected FRBs \citep{LuPiro19_ASKAP_population}, and the horizontal silver line shows the median $\mc{L}(\mbox{ASKAP})=2\times10^{37}\rm\, erg\,s^{-1}\, Gpc^{-3}$. The most active repeaters (R3-like) contribute a similar FRB luminosity density as the least active repeaters (SGR-like or M81-FRB-like), even though their time-averaged luminosities (or repeating rates) differ by many orders of magnitude. }
\label{fig:luminosity-density}
\end{figure}

We show the FRB luminosity density contributed by each of these three subclasses of objects in Fig. \ref{fig:luminosity-density}. We find that the data is consistent with the picture that all three subclasses of FRB sources contribute a similar luminosity density of the order $10^{36}\rm\, erg\, s^{-1}\, Gpc^{-3}$. This is likely a coincidence without a physical reason other than the statistical properties of the FRB population, because the time-averaged luminosity of the most active repeaters (R3-like) is 6-7 orders of magnitude higher than that of the least active repeaters (SGR-like). It can be shown that this coincidence is not due to selection bias. For a fluence-limited survey, the typical energy of detected bursts from a source at distance $D$ scales as $E_{\rm frb}\propto D^2$. Since we are considering the nearest member of each class of sources, the inferred source number density scales as $\nobs\propto D^{-3}$. The inferred luminosity density is given by $\mathcal{L}\simeq \mc{R}_{\rm rep}(>E_{\rm frb}) E_{\rm frb} \nobs$, where $\mc{R}_{\rm rep}(>E_{\rm frb})$ is the cumulative repeating rate above energy $E_{\rm frb}$. For $\mc{R}_{\rm rep}(>E_{\rm frb})\propto E_{\rm frb}^{-\alpha}\propto D^{-2\alpha}$, we obtain $\mathcal{L}\propto D^{-1-2\alpha}$. Since all known repeaters show a decreasing repeating rate with burst energy (and $\alpha\simeq 1$ has been inferred from previous FRB population studies, see footnote\footref{fn:alpha}), one would necessarily conclude that the luminosity density $\mathcal{L}$ inferred from the nearest source rapidly decreases with its distance. Thus, observational bias alone would give rise to $\mc{L}(\mbox{M81-FRB})/\mc{L}(\mbox{R3})\simeq (149\mr{Mpc}/3.6\mr{Mpc})^{1+2\alpha}\ggg 1$, which is in disagreement with eqs. (\ref{eq:mcL_M81FRB}) and (\ref{eq:mcL_R3}) even when Poisson errors are considered. We conclude that the cosmological FRB rate likely has roughly equal contributions from sources across a very broad range of repeating rates. If the luminosity density distribution is indeed flat, suppose we find a single burst at a cosmological distance of $\sim$1 Gpc, it is equally likely that it came from a prolific repeater (from which we expect detections of repetitions) or that it was from a much less active source (from which no detection of repetitions is expected under a reasonable follow-up effort). 

The volumetric rate density of the brightest FRBs with specific energy $E_\nu> 10^{31}\rm\, erg\,Hz^{-1}$ has been well constrained by the ASKAP Fly's Eye survey \citep{Shannon2018}. Note that here we use specific energy $E_\nu$, which is based on the fluence (in units of $\rm\, Jy\, ms$) of each burst. The key advantages of the ASKAP survey are the well-measured fluences of all bursts to a fractional uncertainty of $\sim 20\%$ (thanks to multi-beam detections) and the uniformly sampled focal plane covered by the overlapping beams. Other surveys, by e.g. the CHIME telescope, have widely separated beams (with significant gaps in between adjacent beams) and hence do not accurately measure the fluences of the detected bursts due to their large position errors that are comparable to the beam size. Since the sky is non-uniformly sampled by a highly complex telescope response function that depends on the source position, flux, and spectral shape, it is much more difficult to evaluate the completeness of the resulting sample and then infer the physical rate density of FRBs \citep[this analysis is currently being carried out using comprehensive injection-detection simulations, see][]{chime21_catalog1}.

We use the posterior distribution of the rate density function $\d \mc{R}/\d E_\nu$ from the analysis of the ASKAP sample by \citet{LuPiro19_ASKAP_population} and calculate the luminosity density
\begin{equation}
\begin{split}
    \mc{L}(\mbox{ASKAP}) &=\int_{E_{\nu,\rm min}}^\infty \nu E_\nu {\d \mc{R}\over \d E_\nu} \d E_\nu \\
    &\in (1\times10^{37}, 5\times10^{37}) \rm\, erg\, s^{-1}\, Gpc^{-3}, \mbox{ (90\% CL)}
\end{split}
\end{equation}
where we have adopted $\nu=1.4\rm\, GHz$ as an approximation for the typical spectral width and $E_{\rm \nu,min}=10^{31}\rm\, erg\, Hz^{-1}$ below which the survey loses sensitivity. In fact, the constraints on the faint end of the rate density function by the Galactic FRB200428 \citep{bochenek20_FRB200428, LKZ2020} as well as the M81-FRB (see below) suggest that $\d\mc{R}/\d E_\nu\propto E_\nu^{-\gamma}$ with $\gamma\simeq 2$, so including fainter bursts down to e.g., $10^{26}\rm\, erg\, Hz^{-1}$ will only increase the total luminosity density from $\mc{L}(\mbox{ASKAP})$ by a factor of order unity. 

When Poisson fluctuations are properly taken into account, we find $\mc{L}(\mbox{M81-FRB})/\mc{L}(\rm ASKAP)\gtrsim 1\%$. When other sources with time-averaged luminosities different from the M81-FRB are taken into account, we conclude that FRB sources from old stellar population contribute at least a few percent of the total FRB rate.


Based on Fig. \ref{fig:luminosity-density}, we propose that the time-averaged luminosity function of all observable FRB sources is a power-law of index close to $-1$, i.e.,
\begin{equation}\label{eq:luminosity_function}
    \Phi(\bar{L})\equiv {\d \nobs \over \d\, \mr{ln}\bar{L}} \sim 10^{36}\mr{\,erg \,s^{-1}\, Gpc^{-3}}\, \bar{L}^{-1}.
\end{equation}
Current observations show that the power-law extends from $\bar{L}_{\rm min}\sim 10^{27}\rm\, erg\, s^{-1}$ up to $\bar{L}_{\rm max}\sim 10^{34}\rm\, erg\, s^{-1}$, and this range may be further extended (especially on the lower end) by future observations. The total luminosity density can be obtained by integrating eq. (\ref{eq:luminosity_function}) over the full range of luminosities, and this gives $10^{36}\mr{ln}(\bar{L}_{\rm max}/\bar{L}_{\rm min})\sim 10^{37} \rm\, erg\, s^{-1}\rm\, Gpc^{-3}$, which is in agreement with $\mc{L}(\mbox{ASKAP})$.

The above findings enable us to predict the existence of repeaters with time-averaged luminosities in between SGR and R3, and a way of looking for these sources is to carry out FRB searches in the directions of nearby galaxies, both star-forming and elliptical ones. For instance, let us consider a hypothetical program that monitors $N_{\rm gal}=30$ nearby massive galaxies ($D\lesssim 15\rm\, Mpc$) with an on-source time of $t_{\rm on}=100\rm\, hr$ for each of them. Let us take a luminosity density of $\mc{L}\sim 10^{36}\rm\, erg\,s^{-1}\, Gpc^{-3}$ for the sources that the survey is sensitive to, according to eq. (\ref{eq:luminosity_function}). The average volume shared by each of the galaxies is $\simeq 100\rm\, Mpc^{-3}$, so the total time-averaged FRB luminosity from all surveyed galaxies is $\bar{L}_{\rm tot} \simeq 3\times10^{30}\mr{erg\,s^{-1}}\, (N_{\rm gal}/30)$, and the time-integrated total energy for is $E_{\rm tot}\simeq 1\times10^{36}\rm\,erg\, (N_{\rm gal}/30)(t_{\rm on}/100\mr{hr})$. Suppose the telescope operates at frequency $\nu=1.4\rm\, GHz$ and the fluence-completeness threshold is $F_{\nu\rm,th}=0.1\rm\, Jy\, ms$, so the survey is sensitive to bursts above energy $E_{\rm th} = 4\pi D^2 \nu F_{\nu\rm,th} = 4\times10^{34}\mr{\,erg} (D/15\mr{Mpc})^2 (F_{\nu\rm,th}/0.1\,\mr{Jy\,ms})$. Therefore, the expected number of FRB detections is
\begin{equation}
    N_{\rm frb}\sim {E_{\rm tot}\over E_{\rm th} }\sim 25\, {N_{\rm gal}\over 30}{t_{\rm on}\over 100\mr{hr}} \left(D\over 15\mr{Mpc}\right)^{-2} \left(F_{\nu\rm,th}\over 0.1\,\mr{Jy\,ms}\right)^{-1}.
\end{equation}
This exercise shows the nice property of the (time-averaged) luminosity function in eq. (\ref{eq:luminosity_function}).

We conclude that a survey of nearby galaxies as outlined above is expected to be highly rewarding --- it will allow us to more precisely measure the luminosity function (eq. \ref{eq:luminosity_function}) as well as to accurately determine the fraction of FRB rate contributed by young vs. old stellar populations. More importantly, detecting more local sources and studying their immediate environment will be highly useful for understanding the nature of the emitting objects. Practically, we note that the choice of typical distance $D$ depends on the telescope's field of view (which needs to cover at least the luminous part of the galaxy) and that the number of targets $N_{\rm gal}$ depends on the available time. 




Finally, the above calculations also constrain the FRB rate density function --- the volumetric rate as a function of burst energy --- on the faint end ($E_\nu\lesssim 10^{29}\rm\, erg\,Hz^{-1}$). The rate density function on the bright end ($E_\nu\gtrsim 10^{31}\rm\, erg\,Hz^{-1}$) has been well constrained by ASKAP Fly's Eye survey \citep{Shannon2018}. We find that the volumetric rate above the specific energy of the Galactic FRB200428 inferred from the STARE2 survey is in the range $\mc{R}(E_\nu>1\times10^{26}\rm\, erg\,Hz^{-1})\in (1.8\times10^7, 3.9\times10^8)\rm\, Gpc^{-3}\, yr^{-1}$. The observations of the M81-FRB by CHIME yields a rate constraint at lower energies $\mc{R}(E_\nu>3\times10^{25}\rm\, erg\,Hz^{-1})\in (1.4\times10^8, 8.6\times10^{9})\rm\, Gpc^{-3}\, yr^{-1}$, provided that there are no other sources at the distance of M81 or closer (Fig. \ref{fig:luminosity-density} shows that SGRs in the Milky Way can at most affect this rate density by an order-unity factor). Similarly, the rate density contributed by R3-like sources\footnote{Since there are other FRB sources within the distance of R3 ($D=149\rm\, Mpc$), this constraint should be considered as a lower limit to the total volumetric rate density.} is in the range $\mc{R}(E_\nu>5\times10^{28}\rm\, erg\,Hz^{-1}) \in (4.7\times10^4, 2.8\times10^6)\rm\, Gpc^{-3}\, yr^{-1}$.

\begin{figure}
\centering
\includegraphics[width = 0.48\textwidth]{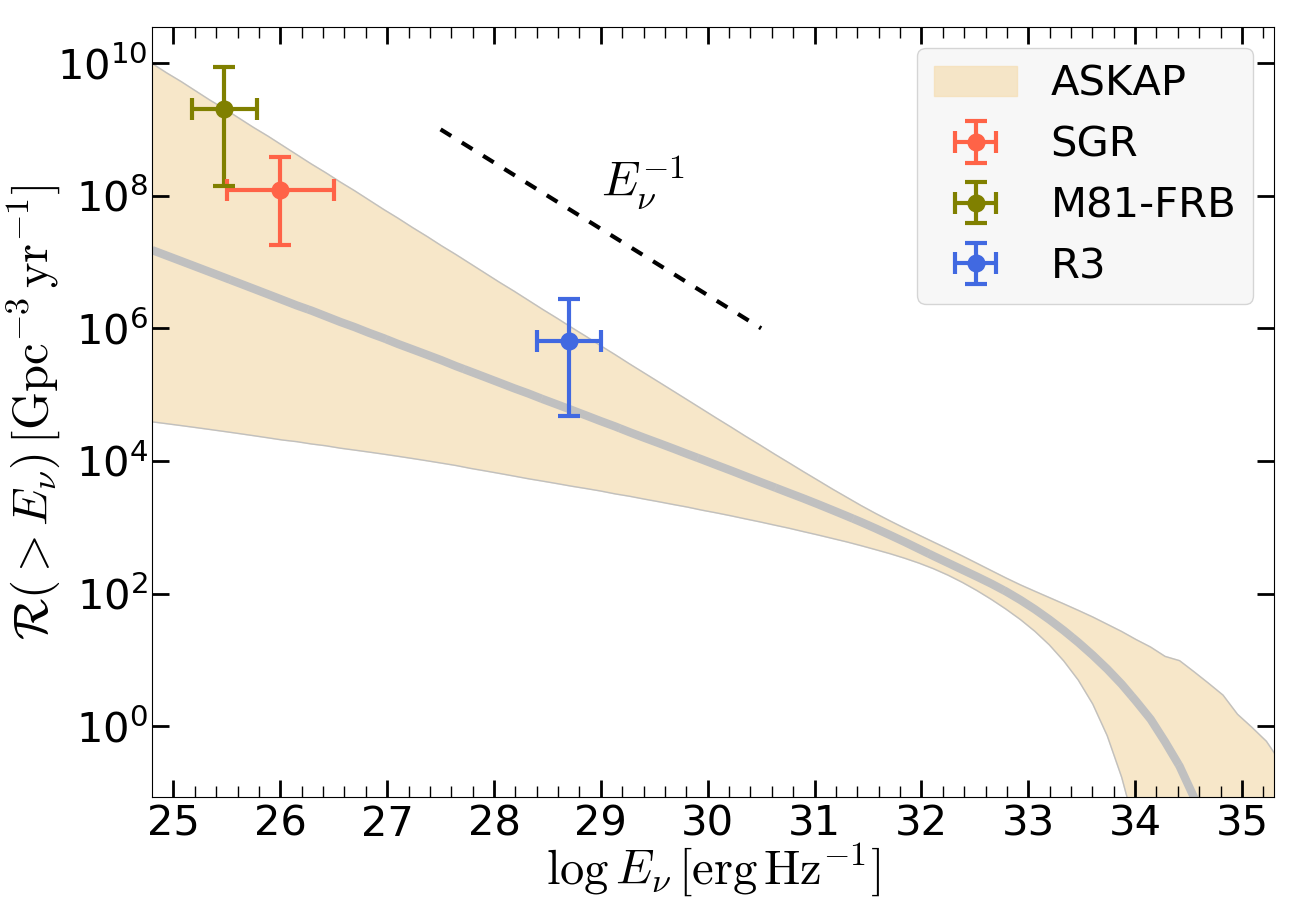}
\caption{Cumulative rate of FRBs as a function of specific energy. The filled circles with errorbars are the estimate from individual sources (SGR, M81-FRB, and R3), which represent their own subclasses. The errors in the vertical direction are due to Poisson fluctuations (90\% CL) and in the horizontal direction are due to imprecise measurements of the energies from individual bursts. The orange shaded band shows the 90\% confidence interval of the Bayesian posterior for the rate density function constrained by the ASKAP sample \citep{LuPiro19_ASKAP_population}, and the silver line shows the median. Since ASKAP only detects the brightest bursts, the constraints on the faint end are very weak. The very nearby FRBs provide better measurements (despite the large Poisson errors) of the faint-end rate than the extrapolation from the ASKAP results. We find the cumulative rate function to be consistent with a power-law $\mc{R}(>E_\nu)\propto E_\nu^{-1}$.
}
\label{fig:rate-density}
\end{figure}

In Fig. \ref{fig:rate-density}, we show the rate density inferred from the three subclasses of FRB sources (SGR, M81-FRB, and R3) as well as the constraints by the ASKAP sample \citep[the posterior is taken from the analysis by][]{LuPiro19_ASKAP_population}. The ASKAP Fly's Eye survey was only sensitive to the brightest bursts in a narrow range of specific energies $E_\nu\in (10^{31}, 10^{33})\rm\, erg\, Hz^{-1}$ \citep[][]{Shannon2018}, so the rate density on the faint end is unconstrained.

With the measurements from nearby events (despite the large Poisson errors), we find that the cumulative rate function is consistent with the following power-law
\begin{equation}
    \mc{R}(>E_\nu)\simeq 3\times 10^{4}\, \mr{Gpc^{-3}\, yr^{-1}}\,  (E_\nu/10^{30}\rm\, erg\, Hz^{-1})^{-1},
\end{equation}
which is applicable in the energy range\footnote{The rate sharply drops above a maximum energy $E_{\nu,\rm max}\sim 10^{33}\rm\, erg\, Hz^{-1}$, but the precise value of the cut-off energy and its physical mechanism are still uncertain. } $E_\nu\in (10^{26}, 10^{33})\rm\, erg\, Hz^{-1}$. More nearby FRBs, from either targeted long-term monitoring of nearby galaxies (as mentioned above) or continuous monitoring of the Galactic plane \citep{connor21_GReX}, are needed to precisely measure the rate function on the faint end and to test whether the power-law extends to even lower energies.

\section{Energetics of the M81-FRB source}\label{sec:M81-FRB-energetics}

In this section, we discuss the nature of the M81-FRB source based on the energetics and its globular cluster association.

The time-averaged isotropic equivalent luminosity $\bar{L}\sim 3\times10^{29}\rm\, erg\, s^{-1}$ can be combined with the lifetime $\tau = 10^4\tau_4\rm\, yr$ (cf. \S \ref{sec:n_star}) to obtain the energy budget of the system
\begin{equation}\label{eq:energy_budget}
    E(\mbox{M81-FRB}) \simeq \bar{L}\tau {f_{\rm b,eff}\over \epsilon_{\rm R}}\simeq 3\times10^{43}\mr{\,erg}\, \tau_4 {f_{\rm b,eff}\over 0.3} \left(\epsilon_{\rm R} \over 10^{-3}\right)^{-1},
\end{equation}
where $f_{\rm b,eff}$ is the beaming factor and $\epsilon_{\rm R}$ is the radio emission efficiency (i.e., the fraction of the total energy release that goes into FRBs).

The only known example of FRB multi-wavelength counterpart is the Galactic magnetar SGR 1935+2154, which showed a radio-to-X-ray fluence ($\nu F_\nu$) ratio of $3\times 10^{-5}$ in FRB 200428 \citep{bochenek20_FRB200428, Li2020, Mereghetti+20}. The fact that many X-ray bursts from this source are not associated with FRBs reduces the time-averaged radio efficiency to be much below $10^{-5}$ \citep{Lin+20}. It is possible that other FRB sources are more efficient at emitting coherent radio waves, so we adopt a rather conservative fiducial value of $\epsilon_{\rm R}=10^{-3} \epsilon_{\rm R,-3}$ in eq. (\ref{eq:energy_budget}). We note that, even for very nearby but extragalactic FRBs, the X-ray counterparts may be extremely difficult to observe with our current instruments (with effective area $\lesssim 10^4\rm\, cm^{2}$). For instance, for a typical FRB energy of $E_{\rm frb}\sim 3\times10^{34}\rm\, erg$, the X-ray fluence at the distance of $D=3.6\rm\, Mpc$ is $\sim 10^{-6}\epsilon_{\rm R, -3}^{-1}\rm\, photon\, cm^{-2}$, where we have taken an energy of $\epsilon_{\rm R}^{-1}E_{\rm frb}$ in the hard X-ray band around $10\rm\, keV$. On the other hand, it is much easier to obtain meaningful constraints in the optical/IR band \citep[due to much lower quantum noise,][]{hardy17_optical_limits, tingay19_TESS_optical_limits, andreoni20_ZTF_optical_limits, chen20_multiwavelength_counterpart, de20_IR_counterparts, kilpatrick21_optical_counterpart}.


If the energy in eq. (\ref{eq:energy_budget}) comes from dissipation of magnetic fields of a NS, we obtain the volume-averaged B-field strength in the NS interior (for a volume of $4\pi R_{\rm ns}^3/3$ and $R_{\rm ns}\simeq 10 \rm\, km$)
\begin{equation}\label{eq:Bfield-limit}
    B\simeq 1.3\times10^{13}\mr{\, G}\, \tau_4^{1/2} \left(f_{\rm b,eff}\over 0.3\right)^{1/2} \left(\epsilon_{\rm R} \over 10^{-3}\right)^{-1/2},
\end{equation}
We conclude that the source is indeed likely a highly magnetized NS, but a Galactic-SGR-like field strength of $\sim 10^{15}\rm\, G$ is not required by current observations.

We also comment on the location of the sudden magnetic dissipations that power FRBs from this source. The magnetic field strength of the FRB waves at a distance $R$ from the NS is given by $B_{\rm frb}(R) =\sqrt{L_{\rm frb}/R^2 c}$, where $L_{\rm frb}=10^{39}L_{\rm frb,39}\rm\, erg\, s^{-1}$ \citep{MajidM81} is the peak isotropic luminosity of the FRB and $c$ is the speed of light. Regardless of the magnetic dissipation mechanism, the magnetospheric B-field strength $B_{\rm mag}$ at radius $R$ is required to be stronger than $B_{\rm frb}$, provided that the energy-conversion processes occur locally. For the case where the emission comes from within the magnetosphere (which is favored by the short variability time, cf. \S \ref{sec:variability-time}), the emitting plasma can only be confined by the magnetosphere if $B_{\rm mag}> B_{\rm frb}$. For the alternative case where the emission comes from a magnetically driven relativistic outflow launched from radius $R$, one also requires that the energy density in the magnetosphere dominates over the FRB wave energy density \citep{yuan20_plasmoid_ejection}. If we take $B_{\rm mag}(R) = B_{\rm d} (R/10\mr{km})^{-3}$, where $B_{\rm d}\simeq 10^{13}\rm\, G$ is the surface dipole field strength, the requirement of $B_{\rm mag}(R)>B_{\rm frb}(R)$ yields $R\lesssim 2\times10^8\mr{\, cm}\, B_{\rm d,13}^{1/2} L_{\rm frb,39}^{-1/4}$. This shows that the burst of energy dissipation must occur in the inner magnetosphere (and this argument alone disfavors millisecond pulsars with $B_{\rm d}\sim 10^8\rm\, G$ as the M81-FRB source).

In the following, we assume that the M81-FRB source is indeed a highly magnetized NS and that the dipolar magnetic field strength $B_{\rm d}$ is comparable to that given by eq. (\ref{eq:Bfield-limit}). Based on these assumptions, we further constrain its properties.

\begin{figure}
\centering
\includegraphics[width = 0.46\textwidth]{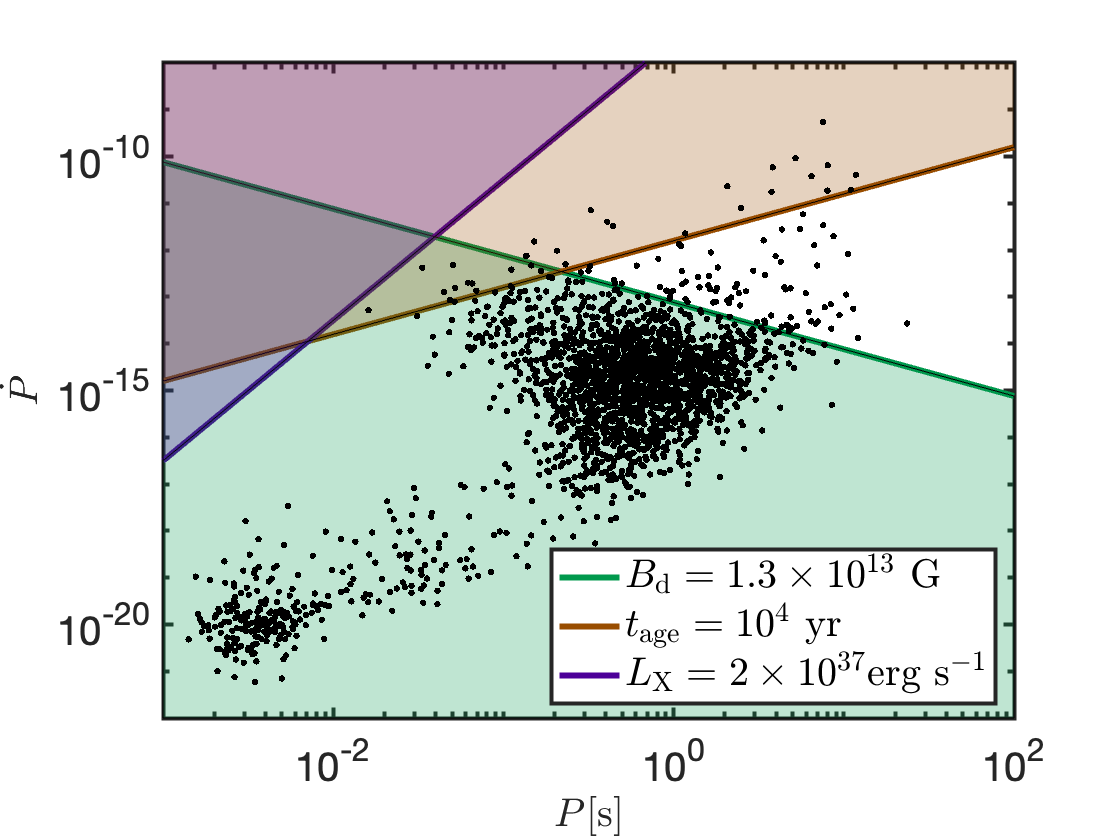}
\caption{Limits on the M81-FRB source in the $P$-$\dot{P}$ diagram (allowed region shown with a white background). The purple-shaded region is excluded by the X-ray upper limit \citep{KirstenM81}. The brown-shaded region indicates where the spindown time of the NS is shorter than $10^4\rm\, yr$, and this is inconsistent with the high number density of M81-FRB-like sources. The green-shaded region shows where the surface dipolar B-field strength $B_{\rm d}$ is smaller than the interior field $B$ as constrained by eq. (\ref{eq:Bfield-limit}). If $B_{\rm d}\ll B$ is allowed, then this constraint will be weaker (the other two constraints based on the X-ray upper limit and age are unaffected). Galactic pulsars taken from the ATNF catalog  \citep{Manchester2005} are shown as black dots.}
\label{fig:PPdot}
\end{figure}

{\it Chandra} observations limit the X-ray luminosity from the M81-FRB source to $L_{\rm X}\lesssim 2\times 10^{37}\mbox{erg s}^{-1}$ \citep{KirstenM81}.
Using the known correlation between X-ray and spindown luminosity in pulsars, $L_{\rm SD} \simeq 10^{39}L_{\rm X,37}^{3/4}\rm\, erg\,s^{-1}$ \citep{Possenti2002}, we obtain an upper limit for the spindown luminosity and this constrains the spin period to be
\begin{equation}
\label{eq:Plim}
    P\gtrsim 0.03\mr{\,s}\, \left(\frac{B_{\rm d}}{10^{13}\rm G}\right)^{1/2} \left(\frac{L_{\rm X}}{2\times 10^{37}\mbox{erg s}^{-1}}\right)^{-3/16},
\end{equation}
where $B_{\rm d}$ is the surface dipolar B-field strength. Another requirement is that the spindown time must be longer than the neutron star's age (which we assume to be comparable to the source lifetime $\tau$), and this gives
\begin{equation}\label{eq:Plim2}
    P\gtrsim 0.2\mr{\,s}\, \tau_4^{1/2} {B_{\rm d}\over 10^{13}\rm\, G}.
\end{equation}
The constraints on the B-field strength, age, and spin period, are summarized in Fig. \ref{fig:PPdot}, alongside Galactic pulsars taken from the ATNF catalog\footnote{\url{https://www.atnf.csiro.au/research/pulsar/psrcat/}} \citep{Manchester2005}. We conclude that a slowly spinning NS ($P\gtrsim 0.2\rm\, s$) with modest B-field strength ($B\gtrsim 10^{13}\rm\, G$), formed in an old stellar environment, is consistent with all the observations of the M81-FRB. We note that a source like the typical Galactic SGRs, although energetically possible, is largely ruled out by the age requirement $\tau\gtrsim 10^4\rm\, yr$ (cf. \S \ref{sec:n_star}).

Alternatively, the FRB emission can in principle come from the NS's spin energy \citep{kremer21_FRB_in_GC}. In this case, the spindown time should be considered as the lifetime of the source and we take the equal sign in eq. (\ref{eq:Plim2}). In order to tap the spin energy, the FRB emission mechanism must exert a negative torque on the star and this has conventionally been taken to be magnetic torque due to the dipole radiation. The brightest giant pulses --- super giant pulses (SGPs) --- from young Galactic pulsars can reach instantaneous luminosities of 10\% of the spindown luminosity $L_{\rm SD}$ \citep[e.g.,][]{Cordes2016}. Analogs of SGPs had been suggested to be responsible for a fraction of FRBs \citep{pen15_giant_pulses_model, Cordes2016, lyutikov16_giant_pulse_model, munoz20_SGP_young_pulsars}. The Crab pulsar is one of the most energetic NSs in the Milky Way, and it emits SGPs with luminosity up to $L\sim 0.1L_{\rm SD}$ and duration of a few $\mu \rm s$ at a rate of $1/300\mr{\,hr}^{-1}$ \citep{bera19_Crab_super_giant_pulses}. It is known that the SGP rate from a given source drops rapidly with luminosity roughly as $\mc{R}(>L)\propto L^{\sim -2}$ \citep{Cordes2016}.

The $\sim 60\mbox{ ns}$ spike measured with DSS-63 \citep{MajidM81} had a peak luminosity of $L_{\rm frb}=2\times 10^{39}\mbox{erg s}^{-1}$ at a central frequency of $2.2\rm\, GHz$. If we take the peak FRB luminosity to be $0.1L_{\rm SD}$, then this leads to an estimate of $L_{\rm SD}$ that is ruled out by the limits on the persistent X-ray emission mentioned above. We note that it is unclear if the SGP luminosity is physically capped at a fraction of $L_{\rm SD}$ or the lack of brighter bursts is simply due to insufficient monitoring time. In any case, since the bursts from the M81-FRB are much more energetic (in terms of both luminosity and energy) than the brightest SGPs from the Crab pulsar and they repeat much more frequently ($\sim$7 in 100 hours), any SGP-based models would generally require a very young and rapidly spinning NS with $L_{\rm SD}\gg 10^{39}\rm\, erg\,s^{-1}$ \citep{munoz20_SGP_young_pulsars}. Therefore, we disfavor the spin-powered scenario.

\section{Short variability time}\label{sec:variability-time}

The observed variability time for the M81-FRB is reported to be about 50 ns \citep{MajidM81}.
In this section, we go through a number of ``external'' medium interactions that might introduce variability to the lightcurve and show that none of these can account for the 50 ns variability of the M81-FRB.

The variability time due to scintillation or plasma-lens is of order the time it takes for a scintillating eddy or lens to move a distance across the sky that is equal to its length, $\ell_s$. For the relative transverse velocity of $v_s$, the variability time is $\sim \ell_s/v_s$. Thus, even for $v_s\sim 10^8$ cm s$^{-1}$, one requires $\ell_s < 5$ cm to provide 50 ns variability. However, considering that the source transverse size must be larger than (50 ns)$\times c\sim 15$m, i.e. much larger than $\ell_s$, these external mechanisms fail to account for the fast variability of the M81-FRB.

Another attractive possibility to consider is that the coherent FRB radiation pulse is fragmented in the longitudinal and transverse directions as a result of its interaction with the plasma in its path due to an instability. This instability, in the nonlinear phase, breaks up the FRB pulse into honeycomb pattern and hence capable of introducing fast variability not present in the original beam. We evaluate whether this mechanism could account for the 50 ns variability. Our discussion follows closely the recent work of \citet{sobacchi21_modulational_instability}. The transverse and longitudinal wavenumbers $(k_\perp, k_\parallel)$ of the fastest growing modes are
\begin{equation}
    c k_\perp \approx a_0 \omega_p,  \quad  c k_\parallel \approx \min\left\{ a_0\omega, \omega_p\right\},
    \label{unstable-k}
\end{equation}
where $\omega_p=(4\pi q^2 n_e/m)^{1/2}$ is electron plasma frequency, $\omega$ is FRB wave frequency, $q$ \& $m$ are electron charge and mass, $n_e$ is electron density, 
\begin{equation}
    a_0 = {qE_0\over mc\omega}
\end{equation}
is a dimensionless strength parameter for the FRB pulse, and $E_0$ is the electric field strength associated with the radio pulse. The growth time for the instability is
\begin{equation}
    \Gamma \approx {a_0^2\omega_p^2\over \omega}.
\end{equation}

The peak radio luminosity for one of the bursts from the M81-FRB was $L_{\rm frb} \sim  10^{39}$ erg s$^{-1}$ \citep{MajidM81}, and thus the strength parameter is
\begin{equation}
    a_0 = 0.05 \, L_{\rm frb,39}^{1/2} R_{13}^{-1} \nu_9^{-1}.
\end{equation}

Let us consider the case where the observed variability time due to the nonlinear honeycomb instability is $\eta$ times the wave period, i.e. $\delta t = \eta (2\pi/\omega)$. Two necessary conditions for the instability to provide this variability follows from eq. (\ref{unstable-k}), 
\begin{equation}
\begin{cases}
    a_0 > \eta^{-1} \quad &\implies \quad R_{13} < 0.05\, \eta L_{\rm frb,39}^{1/2}\nu_9^{-1} \\
    \omega_p >\omega/\eta \quad &\implies \quad n_e > (10^{10} {\rm cm}^{-3})\, \nu_9^2/\eta^2.
\end{cases}
\end{equation}

\begin{figure}
\centering
\includegraphics[width = 0.46\textwidth]{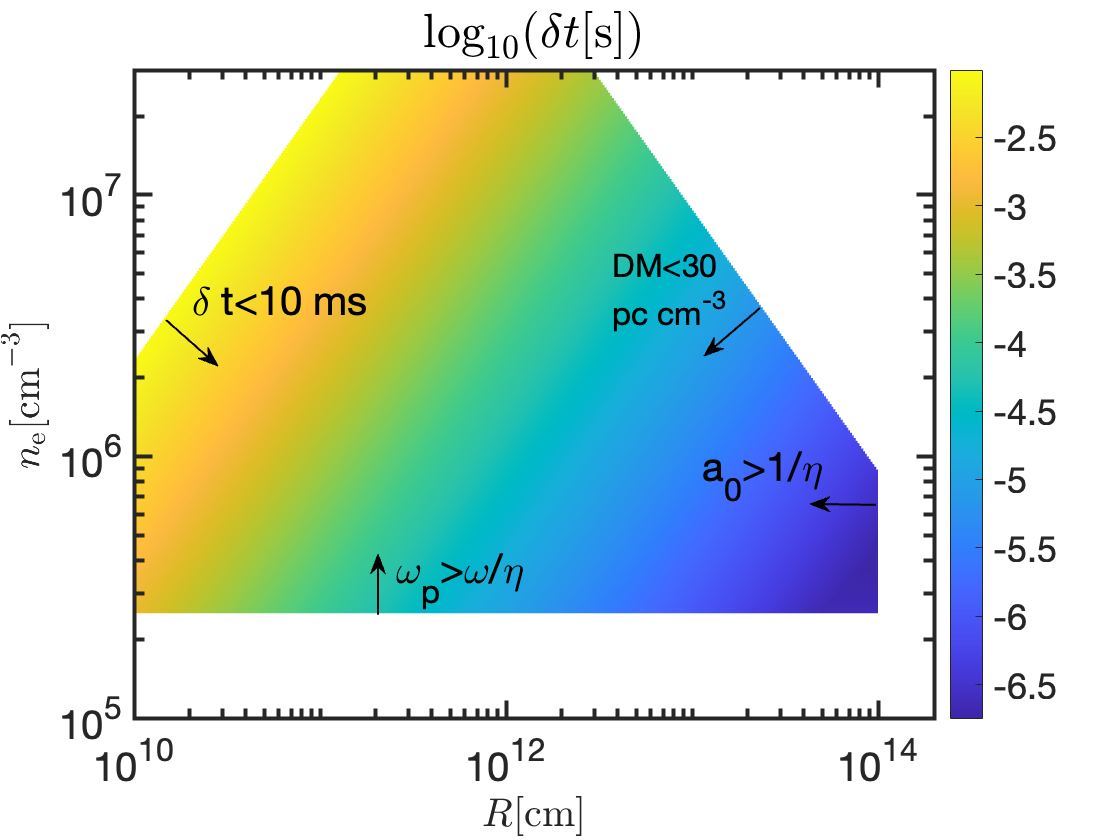}
\includegraphics[width = 0.46\textwidth]{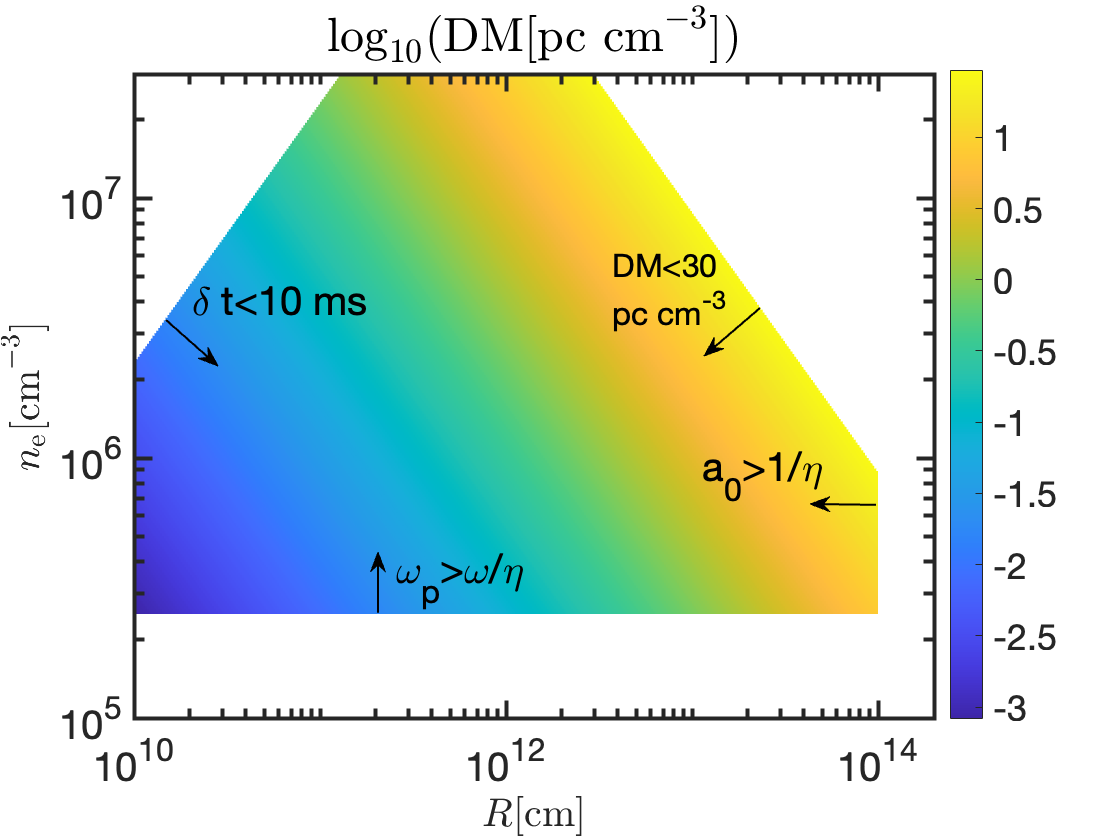}
\caption{Coloured regions depict the parameter space for which the self-modulational instability \citep{sobacchi21_modulational_instability} can produce rapid variability. A lower limit on the variability timescale allowed by the instability is depicted in the top panel and the corresponding contribution of the medium to the DM is shown in the bottom panel. In both panels, the vertical axis is the electron number density of the intervening plasma and the horizontal axis shows the distance to the FRB source. Results are shown for a peak luminosity of $10^{39}\mbox{erg s}^{-1}$ and for an observing frequency of 2 GHz. The shortest variability time due to this instability is $200\rm\, ns$, and in that case, the plasma screen causing this instability would contribute a DM of $5\rm\, pc\, cm^{-3}$. }
\label{fig:selfmodule}
\end{figure}

Wave diffraction due to fragmented wave packets in the transverse direction causes broadening of the honeycomb cells in the longitudinal direction. Since the diffraction angle is given by $\theta_d \sim \lambda k_\perp/2\pi$, the observer receives radiation from an area of radius $R \theta_d$. And this causes the temporal width of the observed pulse to be
\begin{equation}
    \delta t = \max\left\{ 2\pi/(c k_\parallel), R\theta_d^2/c\right\}.
\end{equation}
Using eq. (\ref{unstable-k}), we can show that the scattering time is
\begin{equation}
    \delta t_s = {R\theta_d^2\over c} \approx {a_0^2 \omega_p^2 R\over c\omega^2} \approx {\Gamma R\over c\omega}.
\end{equation}
For the instability to become nonlinear and fragment the wave packet, it requires $\Gamma R/c > 10$. Thus, the condition for producing short time variability by the honeycomb instability is
\begin{equation}
    10 < {\Gamma R\over c} < 2\pi\eta \quad {\rm or} \quad 1.5 < 0.07 {L_{\rm frb,39} n_e\over R_{13} \nu_9^3} < \eta.
\end{equation}
Combining this with the earlier conditions we obtained for $R$ and $n_e$, we find that 
$1.6\times 10^{10} L_{\rm frb,39}^{-1/2} \eta^{-3} < \eta$, or $\eta > 400 L_{\rm frb,39}^{1/8}$. In other words, the shortest variability time this instability could give for the M81-FRB is $\sim 200\rm\, ns$ at 2 GHz, and that is a factor 4 larger than the observed $\delta t$. Furthermore, we note that the plasma screen that causes this variability would contribute $\mr{DM}\sim 5$ pc cm$^{-3}$. The limiting variability timescale and DM given by the plasma screen for which the conditions for the instability are satisfied are shown in Fig. \ref{fig:selfmodule} as a function of the screen density and distance from the source. 

The above considerations suggest that the short variability time of $\sim 50$ ns seen for the M81-FRB is intrinsic, i.e. associated with the mechanism that produced this bright, coherent, radiation. Such a short variability time disfavors \citep{beniamini20_frb_variability} the class of models where the coherent emission is produced by shocks far away from the NS \citep[e.g.,][]{Lyubarsky14, Beloborodov19, Metzger+19, Margalit+20}. On the other hand, the class of emission models where the FRBs are generated from inside the magnetosphere of the NS \citep[e.g.,][]{Kumar+17, LuKumar2018, yang18_curvature_emission} are allowed.

It is easy to show that induced Compton (IC) scattering optical depth of the plasma screen where the honeycomb instability provides the fastest variability is small. This follows from the expression of the IC optical depth \citep[e.g.,][]{kumar20_induced_compton}
\begin{equation}
    \tau_{IC} = {\sigma_T L n_e (c t_{\rm frb})\over 8\pi^2 R^2 m \nu^3},
\end{equation}
which can be rewritten by replacing $n_e$ in terms of the plasma frequency and luminosity in terms of the nonlinear parameter $a_0$ --
\begin{equation}
    \tau_{IC} = {a_0^2 \omega_p^2 t_{\rm frb}\over 3 \nu} = {2\pi\over 3} (\Gamma t_{\rm frb}).
\end{equation}
Therefore, as long as $t_{\rm frb} < R/c$, which is certainly the case for most cases of interest, $\tau_{IC}\ll 1$.

\section{Selection Effects}\label{sec:selection_effects}

At first glance, the association of the M81-FRB with very old stellar environment seems to be in contradiction with the fact many cosmological FRBs with precise localizations are associated with star-forming environment in late-type galaxies \citep[e.g.,][]{heintz20_host_population}. In this section, we point out two selection effects that make the globular cluster-hosted M81-FRB easier to be identified, but we leave it to future works to carefully quantify these selection effects (the conclusions in this work are unaffected).

(1) The excessive DM of the M81-FRB ($\mr{DM}_{\rm tot} \approx 88\rm\, pc\, cm^{-3}$) after subtracting the Galactic ISM ($\rm DM_{\rm MW,ISM}\simeq 35\mbox{--}40 \, pc\, cm^{-3}$) and halo ($\rm DM_{\rm MW,halo}\simeq 30\mbox{--}40\rm\, pc\,cm^{-3}$) contributions is only 10--20$\rm \,pc\, cm^{-3}$ \citep{CHIMEM81, KirstenM81}, and the residual amount is mostly due to the halo of M81. Suppose an identical source is located in the gaseous disk of M81, then the host ISM would have an additional DM contribution of 30--50$\rm\,pc\, cm^{-3}$ if it is in the outskirts and more if closer to the galactic center \citep[][]{Yao2017}. In fact, younger stellar environments tend to contribute larger $\rm DM$'s due to higher gas density and stronger ionizing UV flux. Thus, the hypothetical disk-embedded M81-FRB source would have been mixed into the large pool of unlocalized CHIME sources with $\rm DM_{\rm tot}-DM_{\rm MW,ISM}\sim 100\rm\, pc\,cm^{-3}$. For instance, there are 10 (or 17) sources with $\rm DM_{\rm tot}-DM_{\rm MW,ISM}\lesssim 100\rm\, pc\,cm^{-3}$ (or $\lesssim 120\rm\, pc\, cm^{-3}$) first CHIME catalog \citep[][]{chime21_catalog1}, and there could be others with $\rm DM_{\rm tot}< 1.5\times DM_{\rm MW,ISM}$ that are not included in this catalog.

(2) Another potential bias against the detection of FRBs from young ($\lesssim 100\rm\, Myr$) stellar environment is that the high gas density and strong plasma density fluctuations (due to supernova/UV radiation feedback effects) are likely to cause the radio pulses to be strongly scatter broadened, especially at low frequencies. This may potentially reduce the signal-to-noise ratio and make FRBs more difficult to detect. For instance, the giant pulses from the young pulsar PSR B0540-69 in the star-forming Large Magellanic Cloud have an exponential scattering time of 0.4 ms at 1.4 GHz \citep{johnston03_LMC_pulsar_scattering}; the same pulse at 600 MHz will be broadened to about 17 ms, making it harder to detect.

Additionally, we point out another important but often neglected caveat: the Poisson errors are strongly asymmetric for small number counts. As mentioned in \S \ref{sec:n_star}, from the enclosed volume $V_1$ of the nearest member of a class of sources, one can infer the source number density $\nobs$; the median value of the number density is $\simeq 1/V_1$, but there is non-negligible probability for a much smaller density, e.g., $P(\nobs<0.1 V_1^{-1})\simeq 0.1$ --- the probability only drops linearly with decreasing $\nobs$ in the limit $\nobs\ll V_1^{-1}$ (see eq. \ref{eq:number_density_posterior}).

We conclude that when the above two selection effects as well as Poisson errors are taken into account, the identification of the globular cluster-hosted M81-FRB is not necessarily in contradiction with earlier studies of host galaxies of cosmological FRBs. Nevertheless, we can conservatively draw the conclusion that M81-FRB-like sources from old stellar population contribute at least a few\% of the total FRB rate.

\section{Summary}\label{sec:summary}
The recent discovery of an FRB associated with a globular cluster in the nearby galaxy M81 \citep{CHIMEM81, KirstenM81} clearly demonstrates that there are multiple channels for producing their source objects. Using very general arguments, we show that the repeating activity lifetime of the M81-FRB source is between $10^4$ and $10^6\rm\, yr$ and that its energetics are consistent with a slowly spinning ($P\gtrsim 0.2\rm\, s$) NS with modest magnetic fields $B\gtrsim 10^{13}\rm\, G$. We suggest that this object is formed by a double white-dwarf merger with total mass exceeding the Chandrasekhar mass \citep[such a formation channel has previously be suggested based on hydrodynamic and stellar evolution calculations,][]{shen12_WD_merger_remnants, schwab16_massive_WD_mergers, schwab21_WD_merger_remnant}. If this is true, we predict that M81-FRB-like sources should be found in elliptical galaxies as well as the halo of star-forming galaxies in the future ---  FRBs provide an unprecedented, exciting opportunity to study extragalactic NSs born in old stellar environments.

We estimate the number density of the M81-FRB-like sources to be $n_*\gtrsim 10^{6}\rm\, Gpc^{-3}$, which is much higher than the FRB20180916B- or R3-like sources (whose number density is $\lesssim 10^3\rm\, Gpc^{-3}$). However, the time-averaged FRB luminosity of the M81-FRB is much smaller than R3. We show that the luminosity densities --- defined as the product of the \textit{time-averaged} FRB luminosity and the source number density --- of these two distinct subclasses of sources are likely similar $\mc{L}\sim 10^{36}\rm\, erg\,s^{-1}\, Gpc^{-3}$. We also show that the luminosity density in the radio band contributed by the subclass of sources like the Galactic magnetar SGR 1935+2154 is likely of the same order. This means that the most active (bur rare) R3-like repeaters may contribute similar number of FRBs as the much less active (but abundant) M81-FRB-like or SGR-like sources. When Poisson fluctuations are taken into account, we find that the fractional contribution to the total FRB rate by old stellar populations is at least a few\%. The precise fraction can be constrained by carrying out FRB searches in the directions of nearby galaxies, both star-forming and elliptical ones. This enables us to predict that future interferometric observations of cosmological FRBs will find a fraction of at least a few\% of the sources either in elliptical galaxies and/or at large spatial offsets from the centers of the host galaxies.

Finally, we show that the rapid variability time \citep[$50\rm\, ns$,][]{MajidM81} of the light curves of the M81-FRB is inconsistent with propagation effects far away from the emitting plasma and is hence intrinsic to the emission process (see also \citealt{beniamini20_frb_variability}); this favors the models where the coherent radio waves are generated in a compact region inside the NS magnetosphere \citep{KumarBosnjak2020, LKZ2020}. 

\section*{Acknowledgements}
We thank Eliot Quataert for useful discussion on the evolution of double white dwarf merger remnants and Jonathan Katz for insightful comments on an earlier version of the paper. We also thank the participants of the FRB2021 conference for their questions/comments while the work was presented.
This work has been funded in part by an NSF grant AST-2009619. WL was supported by the David and Ellen Lee Fellowship at Caltech and Lyman Spitzer, Jr. Fellowship at Princeton University. PB was supported by the Gordon and Betty Moore Foundation, Grant GBMF5076.

\section*{Data Availability}
The data underlying this article will be shared on reasonable request to the corresponding author.



\bibliographystyle{mnras}
\bibliography{FRB} 




\appendix


\section{Free Magnetic Energy in neutron star Interior}\label{sec:Bfield-decay-time}
\begin{figure}
\centering
\includegraphics[width = 0.48\textwidth]{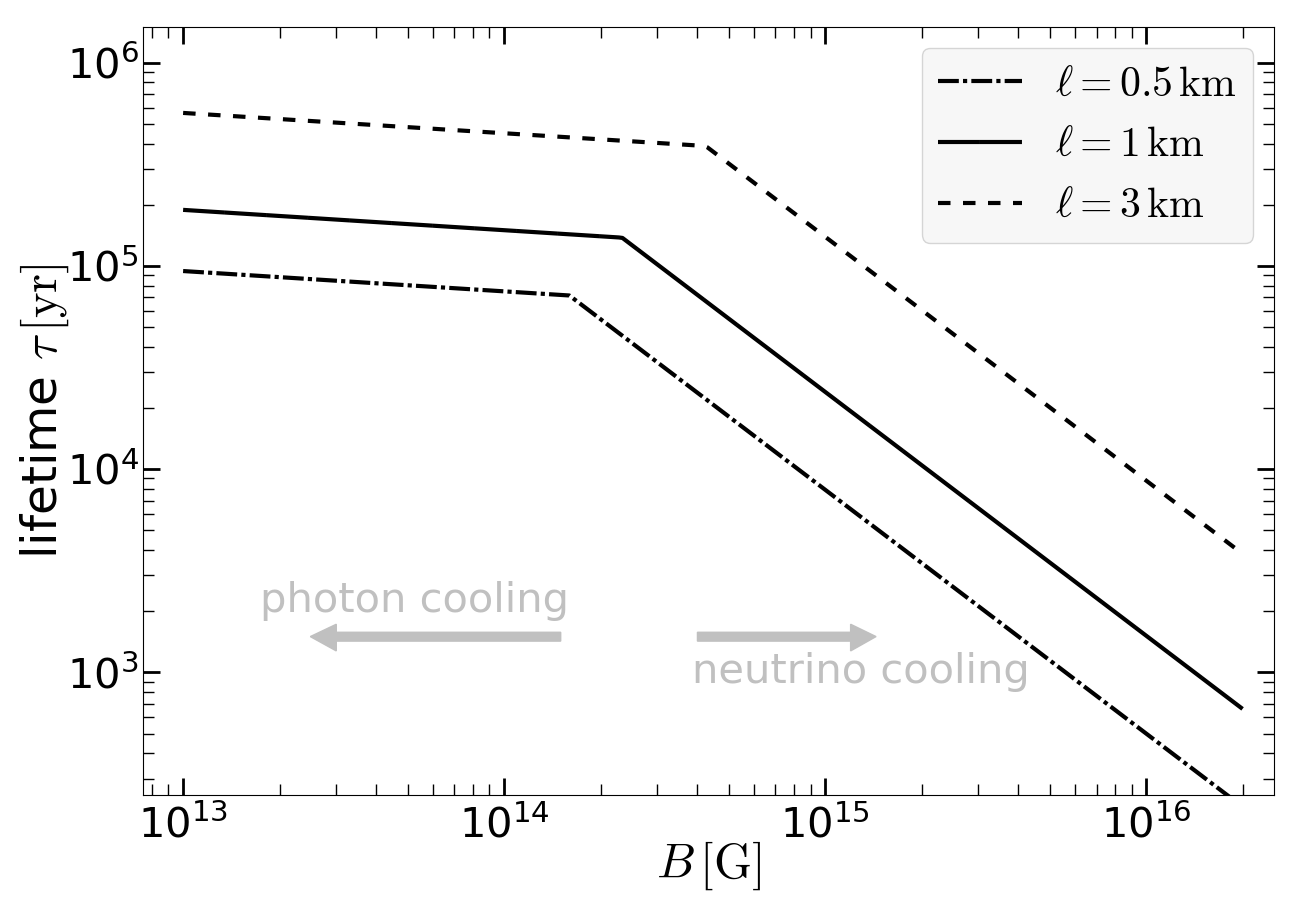}
\caption{Magnetic activity lifetime for neutron stars with various internal B-field strengths, obtained based on ambipolar diffusion timescale \citep{goldreich92_NS_field_decay} for a number of choices of B-field variational length scales $\ell$. In the strong-field regime ($B\gg 3\times 10^{14}\rm\, G$), significant B-field decay occurs while the star is undergoing neutrino cooling (which we assume to be due to modified URCA relactions). In the weak field regime ($B\ll 3\times 10^{14}\rm\, G$), significant B-field decay occurs later when the star has cooled below $\sim10^8\rm\, K$ and the dominant cooling mechanism is thermal photon emission from the surface layer (which we assume to be of pure heavy elements composition). Since the core temperature drops rapidly after photon cooling starts to dominate (roughly $t\sim 10^5\rm\, yr$), the free energy stored in the internal B-fields is expected to be dissipated well within $1\rm\, Myr$, provided that there is no source of replenishment. This conclusion also applies to even lower initial field strengths $B\ll 10^{13}\rm\, G$ for which B-field dissipation is dominated by Hall drift instead of ambipolar diffusion.}
\label{fig:lifetime}
\end{figure}

In this Appendix, we discuss the timescales for the dissipation of free magnetic energy in the NS interior. This sets an upper limit on the source lifetime if FRBs are powered by dissipation of magnetic energy \citep[as suggested by multiwavelength observations of the Galactic FRB as well as many theoretical arguments,][]{LuKumar2018}.

The free magnetic energy stored in strongly magnetized ($B\gtrsim10^{13}\rm\, G$) NSs is slowly dissipated, mainly due to proton-neutron collisions, on the ambipolar diffusion timescale of the solenoidal component \citep{goldreich92_NS_field_decay}
\begin{equation}\label{eq:Bfield-decay}
    \tau_{\rm ambip}\sim 3\times10^5\mr{yr} {\ell_5^2 T_8^2\over B_{14}^2},
\end{equation}
where $\ell=\ell_5\rm\, km$ is the length scale for the B-field variations\footnote{The NS interior likely had initial B-field variations on a wide range of length scales, but the small-scale structures are smoothed out by ambipolar diffusion much faster (eq. \ref{eq:Bfield-decay}) and the largest-scale structures that survive the longest are likely of the order $1\rm\, km$.} in the outer core, $B = 10^{14}B_{\rm 14}\rm\, G$ is the average B-field strength in the outer core, $T=10^8T_8\rm\, K$ is the core temperature (the entire core and the degenerate regions of the crust are nearly isothermal due to high heat conductivity of degenerate matter). The core temperature evolution is governed by the balance between heating and cooling,
\begin{equation}\label{eq:NScooling}
    C_{\rm V}{\d T\over \d t} = L_{\rm B} - L_{\nu} - L_\gamma,
\end{equation}
where $C_{\rm V}\propto T$ is the heat capacity of the core (dominated by degenerate neutrons near the Fermi surface), and $L_{\rm B}\simeq V_{\rm ambip} B^2/(8\pi \tau_{\rm ambip})\propto B^4/T^2$ is the heating power by magnetic field decay ($V_{\rm ambip}\simeq 4\times 10^{18}\rm\, cm^{3}$ being the volume of the outer core), $L_\nu\propto T^8$ \citep[due to modified URCA reactions,][]{shapiro83_compact_objects} and $L_\gamma\propto T^{\sim2.2}$ \citep[near $T\sim 10^8\rm\, K$,][]{page04_NS_cooling} are the cooling luminosities due to neutrino and surface photon emission, respectively.

A generic feature of eq. (\ref{eq:NScooling}) is that the cooling is initially dominated by neutrino emission at very high temperatures ($T\gg 10^8\rm\, K$), and during the neutrino-cooling phase, the temperature follows a power-law function of time $T\simeq 10^{8.5}(t/\mr{kyr})^{-1/6}\,\mr{K}$, where the normalization is approximately correct but subjected to the uncertainties in the core equation of state \citep[see][]{page04_NS_cooling}. In the absence of heating due to B-field decay, photon cooling will dominate over neutrino cooling at a later time $t_\gamma$ when the core has cooled to a characteristic temperature $T_{\gamma}$ such that $L_\nu(T_\gamma)=L_\gamma(T_\gamma)$, and at even later time $t\gg t_\gamma$, the core temperature drops extremely rapidly ($T\propto t^{\sim -5}$) such that the remaining thermal energy is quickly depleted. The radiative cooling rate $L_\gamma$ is affected by the composition of the outer envelope (since lighter elements with lower opacity conduct heat faster than heavier elements) as well as B-field configuration near the NS surface (since strong B-fields only allow heat flow along the field lines) \citep{page04_NS_cooling, BeloborodovLi16}. Based on the analytical expression given by \citet[][their eq. 8]{gudmundsson82_NS_envelope} between surface radiative flux and core temperature based on pure heavy elements composition and vanishing B-fields, we estimate the radiative cooling rate to be $L_\gamma\sim 9\times10^{33}T_{8.5}^{2.2}\mr{\,erg\,s^{-1}}$ (without the general relativistic redshift factor), for a NS of mass $M=1.4M_\odot$ and radius $R=13\rm\, km$. For a neutrino cooling luminosity of $L_\nu\sim 5\times10^{35}T_{8.5}^8\mr{\,erg\,s^{-1}}$ \citep[][their eq. 11.5.24]{shapiro83_compact_objects}, we estimate the critical temperature and time to be $T_\gamma\simeq 1.5\times10^8\rm\, K$ and $t_{\gamma}\simeq 6\times10^4\rm\, yr$.

Heating due to B-field decay can significantly affect the cooling curve of magnetars \citep[e.g.,][]{heyl98_Bfield_decay_heating, vigano13_NS_thermal_evolution, BeloborodovLi16}. If $\tau_{\rm ambip}(T_\gamma)\ll t_{\gamma}$ or $B\gg 3\times10^{14}\rm\, G$, then significant B-field decay occurs at a higher temperature $T_{\rm B,\nu}\gg T_\gamma$ given by $L_{\rm B}(T_{\rm B,\nu})=L_\nu(T_{\rm B,\nu})$. This produces a plateau at temperature $T_{\rm B,\nu}\simeq(2.8\times10^8\mr{\,K}) B_{15}^{0.4}\ell_5^{-0.2}$ and the plateau lasts for $\tau_{\rm ambip}(T_{\rm B,\nu})\simeq(2.4\times10^4\rm\,yr)\, B_{15}^{-1.2}\ell_5^{1.6}$, which we take to be the lifetime for the FRB repeating activity. On the other hand, if $\tau_{\rm ambip}(T_\gamma)\gg t_{\gamma}$ or $B\ll 3\times10^{14}\rm\, G$, B-field decay occurs at temperature $T_{\rm B,\gamma}\ll T_\gamma$ which is given by $L_{\rm B}(T_{\rm B,\gamma})=L_\gamma(T_{\rm B,\gamma})$ since photon cooling dominates. This gives a plateau in the cooling curve at $T_{\rm B,\gamma}\simeq (0.7\times10^8\rm\, K) B_{14}^{0.95}\ell_5^{-0.48}$ and the B-field decay time is given by $\tau_{\rm ambip}(T_{\rm B,\gamma})\simeq (1.5\times10^5\mr{\,yr})\,B_{14}^{-0.1}\ell_5$.

To summarize, based on ambipolar diffusion, we obtain the lifetime for the magnetic activities to be
\begin{equation}\label{eq:lifetime}
    \tau\simeq \mr{min}\left(1.5\times10^5\mr{\,yr}\,B_{14}^{-0.1}\ell_5,\ 2.4\times10^4\rm\,yr\, B_{15}^{-1.2}\ell_5^{1.6}\right),
\end{equation}
which is shown in Fig. \ref{fig:lifetime}. We find that most of the free energy in the internal B-fields is expected to be dissipated well within $1\rm\, Myr$. This is a generic result, because surface photon emission is the dominant cooling mechanism at $t\gtrsim 10^5\rm\, yr$ and after this time the core temperature drops very quickly (and magnetic dissipation rate rapidly increases). This is qualitatively true even in the case where the surface composition has significant light elements or when the modifications to the radiative transfer by strong surface B-fields are taken into account \citep{potekhin03_NS_cooling}. Therefore, we conclude that the lifetime for the FRB-related magnetic activity\footnote{We note that emission powered by the spin of the dipolar field (which does not involve magnetic energy dissipation) may last for much longer than $1\rm\, Myr$. } from a NS is less than $1\rm\, Myr$.


\bsp	
\label{lastpage}
\end{document}